\documentstyle[12pt,epsfig,multirow]{article}

\newcommand{\be}{\begin{eqnarray}}
\newcommand{\ee}{\end{eqnarray}}
\newcommand{\etal}{{\it et al.}}
\def\nue{{\nu_e}}
\def\anue{{\bar\nu_e}}
\def\numu{{\nu_{\mu}}}
\def\anumu{{\bar\nu_{\mu}}}
\def\nutau{{\nu_{\tau}}}

\newcommand{\dm}{\mbox{$\Delta m^2$}}

\newcommand{\chr}{\mbox{$\breve{\rm C}$erenkov~}}

\newcommand{\ms}{\Delta m^2_{21}}
\newcommand{\ma}{\Delta m^2_{31}}

\newcommand{\sss}{\sin^2 \theta_{12}}
\newcommand{\sch}{\sin^2 \theta_{13}}

\newcommand{\sa}{\sin^2 \theta_{23}}

\newcommand{\ses}{\sin^2 \theta_{14}}
\newcommand{\sms}{\sin^2 \theta_{24}}
\newcommand{\sts}{\sin^2 \theta_{34}}
\newcommand{\sess}{\sin^2 \theta_{15}}

\newcommand{\stss}{\sin^2 \theta_{35}}

\newcommand{\smsm}{\sin^2 \theta_{24}^M}
\newcommand{\stsm}{\sin^2 \theta_{34}^M}

\newcommand{\mss}{\Delta m^2_{41}}
\newcommand{\msss}{\Delta m^2_{51}}

\def\ltap{\ \raisebox{-.4ex}{\rlap{$\sim$}} \raisebox{.4ex}{$<$}\ }
\def\gtap{\ \raisebox{-.4ex}{\rlap{$\sim$}} \raisebox{.4ex}{$>$}\ }

\begin{document}

\begin{flushright}
\texttt{HRI-P-07-11-003}\\
\end{flushright}
\bigskip

\begin{center}
{\Large \bf Signature of sterile species in atmospheric 
neutrino data at neutrino telescopes}

\vspace{.5in}

{\bf Sandhya Choubey$^{\star}$}
\vskip .5cm
{\normalsize \it Harish-Chandra Research Institute,} \\
{\normalsize \it Chhatnag Road, Jhunsi, Allahabad  211019, India}
\vskip 1cm
\vskip 2cm

{\bf ABSTRACT}
\end{center}
The MiniBooNE results have still not been able to 
comprehensively rule out the oscillation interpretation of the 
LSND experiment. So far the so-called short 
baseline experiments with energy in the MeV range and 
baseline of few meters have been probing the existence of 
sterile neutrinos.  
We show how signatures of these extra sterile states could be 
obtained in TeV energy range atmospheric 
neutrinos travelling distances of 
thousands of kilometers. 
Atmospheric neutrinos in the TeV range 
would be detected by the upcoming neutrino telescopes. 
Of course vacuum oscillations of these neutrinos would be 
very small. However, 
we show that resonant matter effects 
inside the Earth could enhance 
these very tiny oscillations into 
near-maximal transitions, which should be hard to 
miss.  
We show that imprint of sterile neutrinos 
could be unambiguously 
obtained in this high energy atmospheric neutrino 
event sample. Not only would neutrino telescopes tell the 
presence of sterile neutrinos, it should also be 
possible for them to distinguish between the different possible 
mass and mixing scenarios with additional sterile states.

\vskip 5cm

\noindent $^\star$email: sandhya@mri.ernet.in

\newpage

\section{Introduction}

Last few years have seen a tremendous progress in the field 
of Neutrino Physics, so much so that its fair to say that
neutrinos have now become a pivot in our 
understanding of physics beyond the standard model of 
particle physics. 
The first conclusive prove of the existence of neutrino 
mass and mixing came from the observation of atmospheric 
neutrinos by the SuperKamiokande detector \cite{atm}.
The zenith angle dependent data on atmospheric neutrinos
from this experiment could be explained only if neutrinos 
oscillate with 
$\ma \simeq 2.1\times 10^{-3}$ eV$^2$ and $\sin^22\theta_{23}=1$.  
This was followed by the spectacular results on 
solar neutrinos from the SNO experiment, which 
proved beyond doubt that solar neutrinos do indeed oscillate,
corroborating the observations of all earlier solar 
neutrino experiments, the Homestake, SAGE, GALLEX/GNO, 
Kamiokande and SuperKamiokande \cite{solar}. 
The so-called Large Mixing Angle (LMA) solution 
emerged as the only solution of the solar neutrino deficit 
problem with $\ms = 6\times 10^{-5}$ eV$^2$ and 
$\sss=0.31$ \cite{limits}. The LMA solution was 
confirmed by the KamLAND reactor antineutrino experiment 
\cite{kl} and the combined solar and KamLAND data 
choose $\ms=8\times 10^{-5}$ eV$^2$ and $\sss=0.31$ 
as the best-fit parameter values \cite{limits}. 
The SuperKamiokande atmospheric neutrino results were 
affirmed by two terrestrial accelerator-based experiments -- 
K2K \cite{k2k} and MINOS \cite{minos} and the 
combined atmospheric and accelerator data 
demand $\ma=2.4\times 10^{-3}$ eV$^2$ and 
$\sin^22\theta_{23}=1$ \cite{limits}. Another very important 
result we have is from the reactor antineutrino experiment
Chooz \cite{chooz}. Results of this experiment, 
analyzed along with the other neutrino data impose the 
constraint that $\sch < 0.04$ at $3\sigma$ C.L. \cite{limits}.

The latest addition to the repertoire of 
experimental result on neutrinos comes from the 
MiniBooNE experiment \cite{miniboone}. 
The MiniBooNE experiment was set-up to reconfirm the
positive oscillation  
signal reported by the LSND collaboration \cite{lsnd}, 
which so-far remains the only experiment to have 
seen neutrino oscillations at a frequency which demands
$\Delta m^2$ in the eV$^2$ range. All other short 
baseline experiments \cite{sbl} have been consistent 
with the hypothesis of null oscillations. 
The MiniBooNE experiment also reported to have not 
seen oscillations in the energy regime 
consistent with LSND.  
The extra 
mass squared difference demanded by the LSND signal 
can be accommodated along with solar and atmospheric 
neutrino results only if there were sterile neutrinos. 
The most economical scenario comes from adding just one 
extra sterile neutrino, giving the so-called 2+2 
and 3+1 neutrino mass schemes \cite{sterileold}.
It has been shown that the 2+2 spectrum is strongly 
disfavored from the solar and atmospheric neutrino data.
The 3+1 scheme on the other hand suffers from a strong 
tension between the positive signal at LSND and null 
signal in all other short baseline experiments. 
The addition of the MiniBooNE results 
puts even stronger constraints 
on the 3+1 picture, disfavoring it at a very high C.L. 
\cite{maltonischwetz}. Adding two extra sterile neutrino 
would give us the so-called 3+2 neutrino mass spectrum 
\cite{threeplustwo,sn3p2,ws3p2}. This picture 
interestingly gives a reasonable explanation of all neutrino 
oscillation data including LSND and MiniBooNE, if 
CP violation is allowed \cite{maltonischwetz}. In 
\cite{karagiorgi} the author find a very good fit to 
world neutrino data for a CP conserving 3+2 mass 
spectrum as well. 

The situation concerning sterile neutrinos therefore seems to be 
far from settled. MiniBooNE was especially designed to 
confirm or refute the LSND signal and they have reported 
to have contradicted the LSND claim of positive oscillation signal.
However, their first data set is with neutrinos while LSND had 
seen oscillations of antineutrinos. In addition the entire 
event sample of MiniBooNE is not yet fully understood.  
They have seen excess 
electron events in their low energy sample, which still
remains unexplained. It is hoped that this systematic excess 
of electron events seen in the experiment will eventually be 
explained. MiniBooNE is also now running in the antineutrino 
channel and results from this data set might 
settle the issue regarding the mismatch between the LSND and 
MiniBooNE results. 

Resolution of this perplexing issue could 
also come from other kind of experiments. 
Presence of sterile neutrinos would lead to 
distinctive features in the resultant supernova neutrino
signal in terrestrial detectors such as future 
megaton water \chr detectors and 
neutrino telescopes like IceCube \cite{sn3p2}. 
Very recently 
it has been shown that the planned and up-coming 
next generation reactor neutrino 
experiments such as Double Chooz, Daya Bay, Angra, and RENO,
which are being built to probe the mixing angle $\theta_{13}$, 
could also check for the existence of sterile neutrinos through 
combination of data from the near and far detectors \cite{dcsterile}.
Not only should it be possible to cross-check the 3+2 
neutrino mass scheme at these experiments, we should also be 
able put limits on the mixing angles involving  
sterile neutrinos \cite{dcsterile}. 
Possibility of observing sterile neutrinos in the 
upcoming accelerator-based long baseline experiments 
was studied very recently in \cite{doninisterile} for 
the conventional CNGS experiment, and in 
\cite{shamayita} for the future neutrino factory.
In another recent 
paper we have shown that the existence of sterile neutrinos 
could in principle 
also be probed in the ultra high energy 
neutrino signal in the neutrino telescopes \cite{nutelsterile}. 

In this paper we show how the data from very high energy 
atmospheric neutrinos in neutrino telescopes such as 
IceCube \cite{icecube}, Km3Net \cite{km3net}, NEMO \cite{nemo}
and NESTOR \cite{nestor}
could be used to check if sterile neutrinos indeed exist.
The neutrino telescopes will have energy threshold of about 
100 GeV and 
are designed to observe 
ultra high energy neutrinos. The very high energy 
atmospheric neutrinos with energy in the range $10^{-1}-10^4$ TeV
will also be observed in these detectors and they will in 
principle constitute the ``background'' for the ultra 
high energy neutrino ``signal''. However, this atmospheric 
neutrino ``background'' in km$^3$ neutrino telescopes 
will be sizable and can hence be used to provide 
crucial information on some physics issues. 
The AMANDA experiment has already observed the 
high energy atmospheric neutrinos, and 
the observed flux is reported to be consistent with 
the theoretical predictions \cite{amandaatm}.
In 10 years of operation, IceCube will be 
able to collect $7\times 10^5$ atmospheric 
muon neutrino events \cite{nutelnsi}. With such a 
huge data sample, it was shown    
in \cite{nutelnsi} that the atmospheric 
neutrino events in IceCube could be used to put severe 
constraints on non-standard physics. 
Feasibility studies of constraining non-standard physics
in ANTARES was performed in \cite{nutelantares}.
We will show that for the mass squared difference needed to 
explain LSND, we expect near-resonant matter effects between the 
active and sterile neutrino states inside the Earth. 
This leads to drastic changes in the expected 
flux at the detector. This change is both energy as well as 
zenith angle dependent and should provide foolproof 
signal for the existence of sterile neutrinos in this 
mass regime.

The paper is organized as follows. In Section 2 we 
discuss the enhancement of neutrino mixing and 
oscillations from the  
matter effects due to the extra sterile states. 
The numerically calculated exact 
oscillation probabilities for the PREM matter density 
profile of the Earth is presented in Section 3 for the 
simpler (though disfavored) 3+1 spectrum. 
We reiterate that we 
present results for this case purely for illustration 
only. With the 3+1 exemplary case, we hope to 
highlight some of the features of the active-sterile resonant 
matter effects.
In section 4 we show the oscillation probabilities for the 
more realistic 3+2 neutrino mass spectrum. We 
make some comments in Section 5 
on the detection of the atmospheric 
neutrinos in neutrino telescopes and possible signatures of 
sterile neutrinos in the data sample.  
We end in Section 6 with discussions and conclusions.

\section{Neutrino Mixing in Matter with Sterile Neutrinos}

Neutrinos would undergo 
maximum flavor conversion in vacuum when the oscillatory term  
\be
\sin^2\bigg(\frac{\Delta m^2_{ji} L}{4E}\bigg) = 1~,
\ee
where $L$ and $E$ are the 
distance travelled by the neutrinos and $\Delta m^2_{ji} = m_j^2-m_i^2$.
This happens when 
their energy corresponds 
to the value 
\be
E~({\rm TeV}) &=& 0.81\times 10^{-3}\, \bigg(\Delta m^2_{ji}~({\rm eV}^2)\bigg)\,\bigg(L~({\rm km})\bigg)~,
\nonumber\\
&=& 8.1\times \bigg(\frac{\Delta m^2_{ji}}{1~{\rm~eV}^2}\bigg)\times
\bigg(\frac{L}{10,000~{\rm~km}}\bigg)~.
\ee
Thus we see that if sterile neutrinos are mixed with the active 
ones with $\Delta m^2_{ji}\sim $ eV$^2$, we expect to see maximum 
flavor conversions for neutrinos with energies in the range of 
a few TeV, if they are traveling over distances in the 
range of 10,000 km. 
High energy atmospheric neutrinos travel distances 
of this order to reach the neutrino telescopes. They 
would therefore encounter these flavor oscillations. 
The amplitude and hence the 
extent of the oscillations is of course determined 
by the corresponding mixing angle, which for sterile 
neutrinos are usually constrained to be extremely small
\footnote{See \cite{nutelsterile} for consequences of 
large active-sterile mixing for ultra high energy 
neutrino signal in the neutrino telescopes.}. 
The combined errors coming from the uncertainties 
in the predicted high energy atmospheric neutrino 
fluxes and the experimental uncertainties, could 
threaten to wash out these oscillations driven by tiny 
mixing angles. 

However, atmospheric neutrino travel through the 
matter before they reach the detector and this 
could produce drastic changes in the amplitude of 
the active-sterile oscillations\footnote{Matter effects
for very high energy neutrinos 
due to presence of sterile neutrinos has been discussed before 
in some form in \cite{nicolaidis,yasuda3p1matter,nunokawa}. }.
In fact,
the focal point of this paper is the very large matter effects 
which the neutrinos pick as they move inside the Earth's matter.
In presence of matter the neutrino mass squared matrix changes to
\cite{msw1,msw2,msw3}
\be
M_F^2 = U{\cal M}U^\dagger + {\cal A}
\ee
where $U$ is the unitary mixing matrix relating the
mass eigenstates to the flavor eigenstates, 
\be
{\cal M} = Diag(m_1^2,m_2^2,m_3^2,m_4^2)~~{\rm or}~~
Diag(m_1^2,m_2^2,m_3^2,m_4^2,m_5^2)~,
\ee
\be
{\cal A}=Diag(A_{CC},0,0,A_{NC})~~ {\rm or}~~
Diag(A_{CC},0,0,A_{NC},A_{NC})\,,
\label{eq:mat}
\ee
for the 3+1 and 3+2 neutrino mass spectrum respectively, 
where
\be
A_{CC}=\pm 2\sqrt{2}G_{F} \rho N_A Y_{e}E\,, \label{eq:5ccmat}
\ee
\be
A_{NC}=\pm \sqrt{2}G_{F}\rho N_A (1-Y_{e})E \,.\label{eq:A2}
\ee
Here 
the quantities $A_{CC}$ 
and $-A_{NC}$ are the matter induced charged current and 
neutral 
current potentials respectively, and given in terms of  
the Fermi constant 
$G_F$, matter density $\rho$, Avagadro number $N_A$, 
electron fraction $Y_e$ and energy of the neutrino $E$. The 
``$+$'' (``$-$'') sign in Eqs. (\ref{eq:5ccmat})
and (\ref{eq:A2}) corresponds to neutrinos (antineutrinos).
In the above equations we have re-casted the mass matrix 
in such a way that the 
neutral current component $-A_{NC}$, which is negative and which 
appears for all the 
three active flavors, is filtered out from the first three diagonal 
terms and hence it stays back as positive 
$A_{NC}$ for the sterile state(s), since they do not 
have any weak interactions. 
Presence of matter 
dependent terms in the mass matrix 
modify the mass squared differences and 
mixing angles of the neutrinos in matter and  
these quantities are given respectively as 
\be
(\Delta m^2_{ji})^M &=& 
\sqrt{(\Delta m^2_{ji}\cos2\theta_{ij} - A_M)^2
+(\Delta m^2_{ji}\sin2\theta_{ij})^2}\,, 
\\
\sin 2\theta_{ij}^M &=& \sin 2\theta_{ij}
\frac{\Delta m^2_{ji}}{(\Delta m^2_{ji})^M}\,.
\ee
In the above we have assumed that 
only two neutrino
states are predominantly involved and depending 
on which neutrino states these are, $A_M$ could be 
$A_{CC}$, $A_{NC}$ or 
$A_{CC}-A_{NC}$. 
In this  approximation 
the condition for $\nu_e\rightarrow \nu_{\mu}$ 
or $\nu_e \rightarrow \nu_{\tau}$
resonant 
transition when $\sin^22\theta_{ij}^M =1$, is given by
\be
A_{CC} = \Delta m^2_{ji} \cos2\theta_{ij}~,
\label{eq:emt}
\ee
where $\Delta m^2_{ji}=m_j^2-m_i^2$ 
is the mass squared difference between the 
two involved states and $\theta_{ij}$ is the corresponding mixing 
angle. For $\nu_\mu \rightarrow \nu_{s}$ ($\nu_s$ is a sterile state) or 
$\nu_\tau \rightarrow \nu_{s}$ resonant 
transition the condition is 
\be
A_{NC} = -\Delta m^2_{ji} \cos2\theta_{ij}~,
\label{eq:smt}
\ee
while for $\nu_e\rightarrow \nu_s$ the resonance condition is
\be
A_{CC} - A_{NC} = \Delta m^2_{ji} \cos2\theta_{ij}~.
\label{eq:es}
\ee
Since $A_{CC}$ and $A_{NC}$ are both positive 
for neutrinos and both negative for antineutrinos, 
and since $Y_e \simeq 0.5$ giving   
$A_{CC} \simeq 2A_{NC}$ for Earth matter, 
this resonance condition 
(\ref{eq:es}) is satisfied for neutrinos 
when $\Delta m_{ji}^2 > 0$ 
and for antineutrinos when $\Delta m_{ji}^2 < 0$. 
On the other hand, the $\nu_\mu \rightarrow \nu_{s}$ or 
$\nu_\tau \rightarrow \nu_{s}$ resonance condition will be 
satisfied for  neutrinos (antineutrinos)
when $\Delta m_{ji}^2 < 0$ ($\Delta m_{ji}^2 > 0$).
The $\nu_e \rightarrow \nu_{\mu}$ 
or $\nu_e \rightarrow \nu_{\tau}$ resonance will happen 
in the neutrino (antineutrino) channel when 
$\Delta m_{ji}^2 > 0$ ($\Delta m_{ji}^2 < 0$). 
Since we have both neutrinos and antineutrinos coming from the 
atmosphere, we could have resonance in either the 
neutrino or the antineutrino channel for a given 
$sgn(\Delta m_{ji}^2)$, if any of the above conditions 
are satisfied. 
 
All results presented in this paper are generated 
by exactly solving the full set of evolution equations for the neutrinos 
travelling through Earth matter parameterized by the PREM 
density profile \cite{prem}.   However, just for simplicity 
let us for the moment assume that 
the neutrino travel through constant density matter inside
the Earth. Assuming that $Y_e=0.5$, 
the very high energy atmospheric neutrinos 
travelling through the Earth would pick matter 
potential of
\be
A_{CC} &=& 1.907~({\rm eV^2})\times \bigg(\frac{\rho}{5.0{~\rm gm/cc}}\bigg)
\bigg(\frac{E}{5.0{~\rm TeV}}\bigg)
~,
\\
A_{NC} &=& 0.954~({\rm eV^2})\times \bigg(\frac{\rho}{5.0{~\rm gm/cc}}\bigg)
\bigg(\frac{E}{5.0{~\rm TeV}}\bigg)
~,
\\
A_{CC} - A_{NC} &=& 0.953~({\rm eV^2})\times 
\bigg(\frac{\rho}{5.0{~\rm gm/cc}}\bigg)
\bigg(\frac{E}{5.0{~\rm TeV}}\bigg)~.
\ee
Therefore, we see that the resonance conditions given by Eqs.
(\ref{eq:emt}), (\ref{eq:smt}), and (\ref{eq:es}) can be 
easily satisfied for $\Delta m^2_{4i}$ and $\Delta m^2_{5i}$
relevant for LSND and MiniBooNE, 
where $i=1,2,3$.
We could also turn around these equations to give us the 
resonance energy as
\be
E_{res}^{\nu_e\rightarrow\nu_a}({\rm TeV}) 
&=& \frac{\Delta m_{4i}^2\cos2\theta_{i4}}
{0.076\times \bigg({\rm \rho(gm/cc)}\bigg)}~,
\\
E_{res}^{\nu_a\rightarrow\nu_s}({\rm TeV}) 
&=& \frac{\Delta m_{4i}^2\cos2\theta_{i4}}
{0.038\times \bigg({\rm \rho(gm/cc)}\bigg)}~,
\\
E_{res}^{\nu_e\rightarrow\nu_s}({\rm TeV}) 
&=& \frac{\Delta m_{4i}^2\cos2\theta_{i4}}
{0.038\times \bigg({\rm \rho(gm/cc)}\bigg)}
~,
\ee
where $a$ in the above equations refer to either the $\mu$ or 
$\tau$ flavor. 
For the $\Delta m_{5i}^2$ case we have the same relations with 
$\Delta m_{4i}^2$ replaced by $\Delta m_{5i}^2$. 
Therefore, for neutrinos crossing the core (mantle only) 
of the Earth for which the average matter density is $\sim 8$ gm/cc 
($\sim 5$ gm/cc),
we expect resonance for $E\sim 3$ GeV 
($\sim 5$ GeV) if we assume $\Delta m_{4i}^2$
and $\Delta m_{5i}^2$ to be about 1 eV$^2$. 
We note that the energy at which we 
expect to see $\nu_e\rightarrow\nu_s$ resonance is the 
same as the one where we are expecting to get 
$\nu_a\rightarrow\nu_s$ resonance. 
On the other hand the energy at which 
we will get $\nu_e\rightarrow\nu_a$ resonance will be 
lower by a factor of about 2.
We get similar expressions also for $\Delta m^2_{5i}$ induced 
resonances. Note that even though we have given the 
discussion for the 
$\dm\sim$ eV$^2$ driven 
$\nu_e\rightarrow\nu_a$ resonance for completeness,
this resonance never happens in the 3+1 or 3+2 scenario 
inside the Earth since the $\dm$ involved between the $\nu_e$ and 
$\nu_a$ states are the ones needed to explain 
the solar and atmospheric neutrino data and hence definitely not 
of the eV$^2$ scale that we are interested in. The mass 
eigenstates $\nu_4$ and $\nu_5$ are
predominantly composed of 
the sterile components and hence the $\mss$ and $\msss$ 
mass squared difference drive the active-sterile 
resonances only. We remind the reader of the 
well known fact that when the resonance condition is satisfied, 
the corresponding mixing angle, even if it was very small in 
vacuum, becomes maximal in matter. Thus the amplitude factor in 
the oscillation probability also becomes maximal.

The 
oscillation probabilities in matter are given by the 
most general expression 
\be
P_{\beta\gamma} (L) =& \delta_{\beta\gamma}& - 4 \sum_{j>1} \Re
\left(U^M_{\beta i} U^{M^\star}_{\gamma i} U^{M^\star}_{\beta j}
U^M_{\gamma j}\right) \sin^2 {(\Delta m^2_{ij})^M L \over 4E}
\nonumber \\
&&+ 2 \sum_{j>1} \Im \left(U^M_{\beta i} U^{M^\star}_{\gamma i}
U^{M^\star}_{\beta j} U^M_{\gamma j}\right) \sin {(\Delta m_{ij}^2)^M L
\over 2E},
\label{eq:pr}
\ee
where 
$(\Delta m^2_{ij})^M$ and $U^M$ 
are respectively the modified mass squared 
difference and 
mixing matrix in matter.  
The mixing matrix is parameterized in terms
of the mixing angles (in matter). For the 3+1 case we will have 
6 mixing angles, while for the 3+2 scenario $U^M$ is given 
in terms of 10 angles. In fact, the most general form of 
the mixing matrix $U^M$ is complex and this CP dependence 
is probed through the last term in Eq. (\ref{eq:pr}).  
But for simplicity, we will put all CP violating phases 
in $U^M$ to zero and 
hence the last 
term in Eq. (\ref{eq:pr})
goes to zero. Also for TeV energy neutrinos, the 
oscillations induced by $\ms$ and $\ma$ are negligible
and we will get contributions 
from mainly the oscillatory terms corresponding to the mass squared 
difference associated with the sterile states. We reiterate that 
each term in the oscillation probability contains a product of the
mass squared dependent oscillatory term and the mixing angle 
dependent term. Therefore, to achieve maximal oscillations in 
Earth matter, it is not enough to satisfy only the condition 
of resonance where the mixing angle becomes maximal. We should 
should simultaneously have the peak of the oscillatory term 
\cite{maxmatter}. Thus one obtains maximal oscillations when 
the condition 
\be
\rho L~({\rm km~gm/cc}) = \frac{33.55\times 10^3}{\tan2\theta_{ij}}~,
\label{eq:maxoscgen}
\ee
is satisfied. We have assumed a constant density for the Earth matter, 
$L$ is the baseline where maximal oscillations happen and 
$\rho$ is the corresponding average density. 
We will discuss this issue in greater details in 
the following section.

\section{Neutrino Oscillations with 3+1 Mass Spectrum}

\begin{figure}
\begin{center}
\includegraphics[width=10.0cm, height=7cm, angle=0]{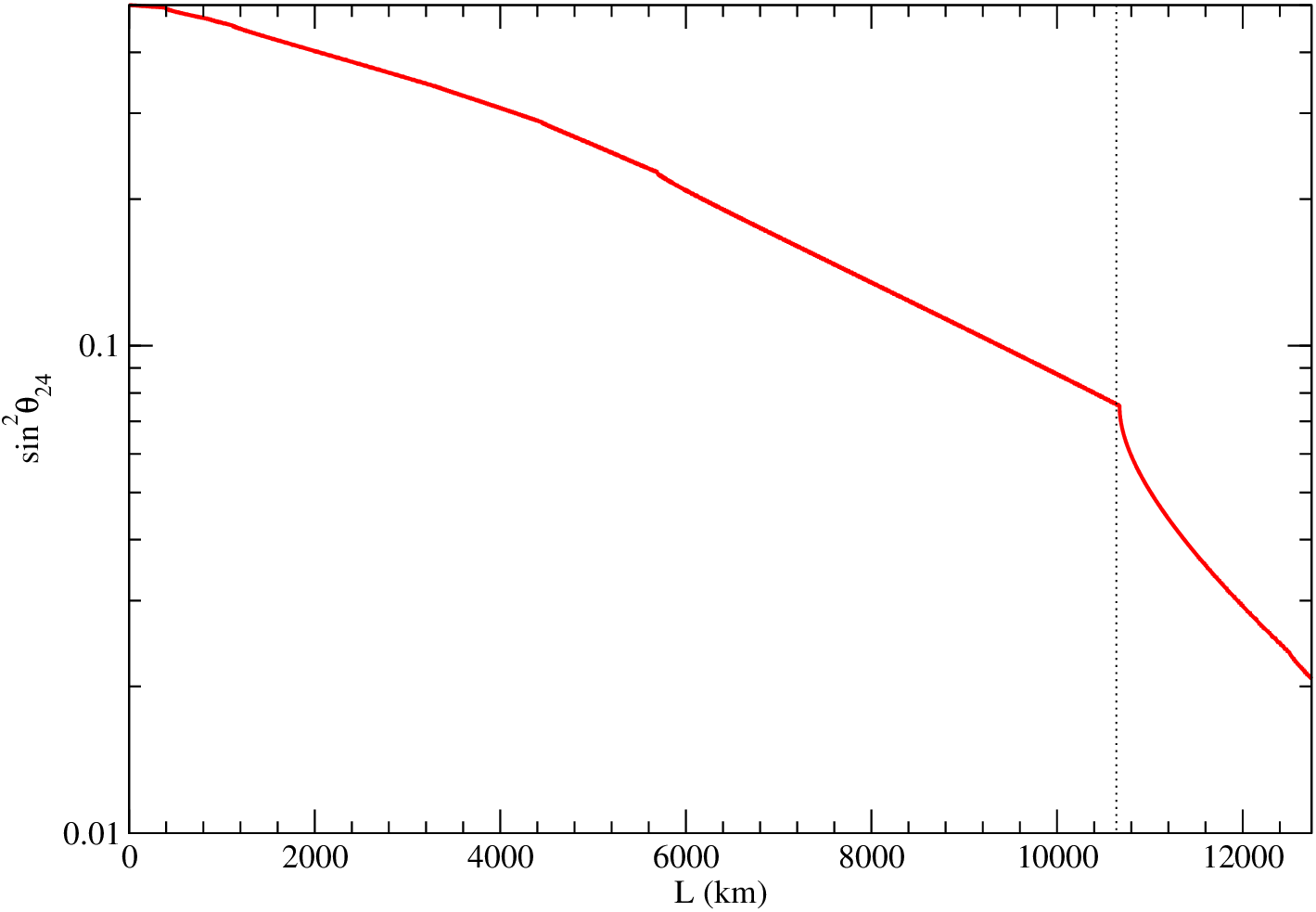}
\caption{\label{fig:maxosc}
Values of $\sms$ at which we have maximal oscillations 
as a function of the distance $L$ travelled inside 
Earth. We have assumed $\ses=0$ and $\sts=0$ and the 3+1 mass 
spectrum. The dashed line shows the boundary between the 
mantle and core of the Earth.  
}
\end{center}
\end{figure}

\begin{figure}[t]
\begin{center}
\includegraphics[width=16.0cm, height=9cm, angle=0]{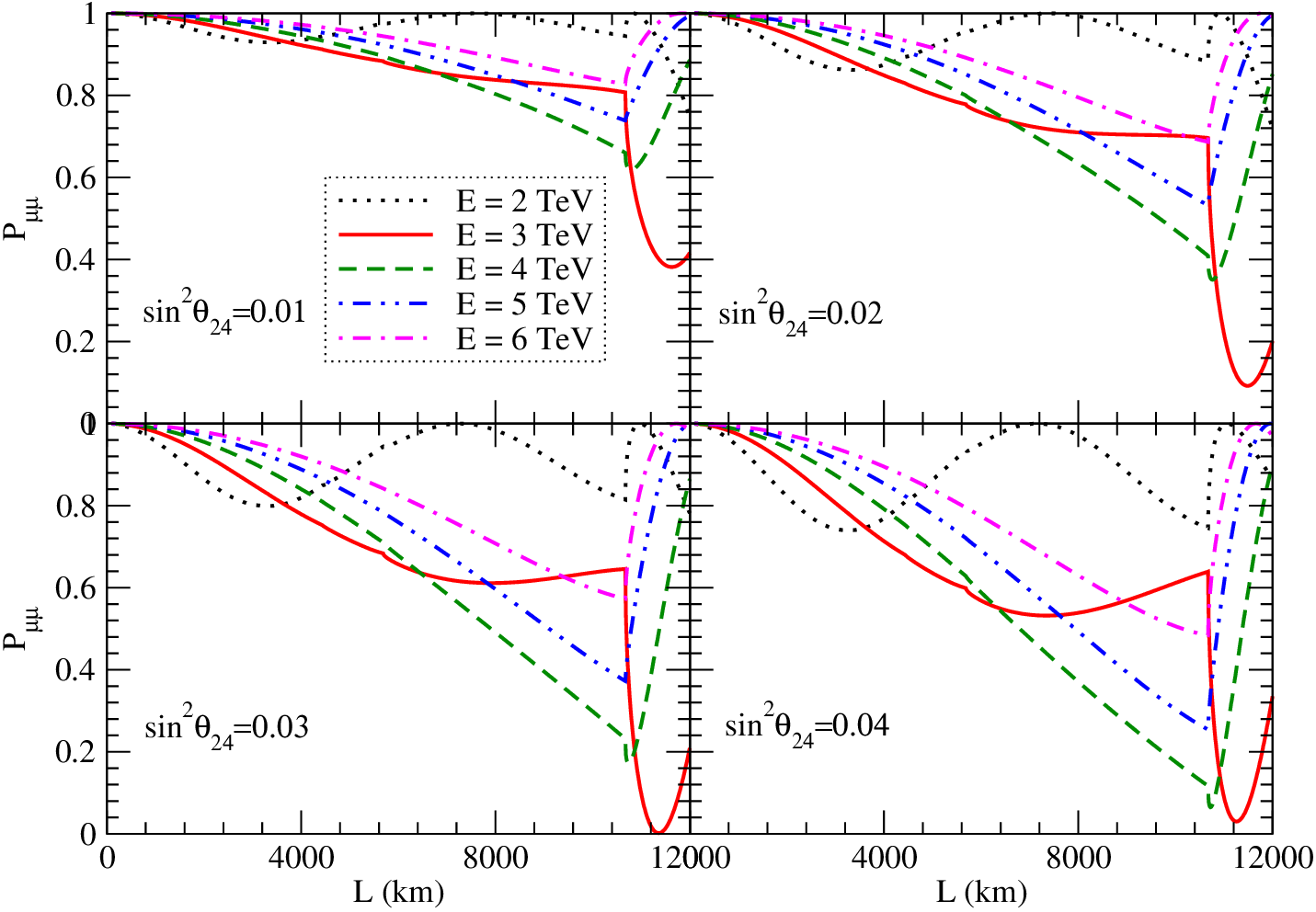}
\caption{\label{fig:LfixedE}
The survival probability $P_{\mu\mu}$ as a function of 
$L$ using the PREM profile for the Earth density. 
The different line types correspond to different fixed 
values of $E$ and each panel shows the results for different 
fixed values of $\sms$ given in the panels. 
We have assumed the 3+1 mass 
spectrum and taken $|\mss|=1$ eV$^2$, $\ses=0$ and $\sts=0$. 
The probability corresponds to neutrinos for $\mss < 0$ and 
to antineutrinos for $\mss > 0$.
}
\end{center}
\end{figure}

\begin{figure}[t]
\begin{center}
\includegraphics[width=16.0cm, height=9cm, angle=0]{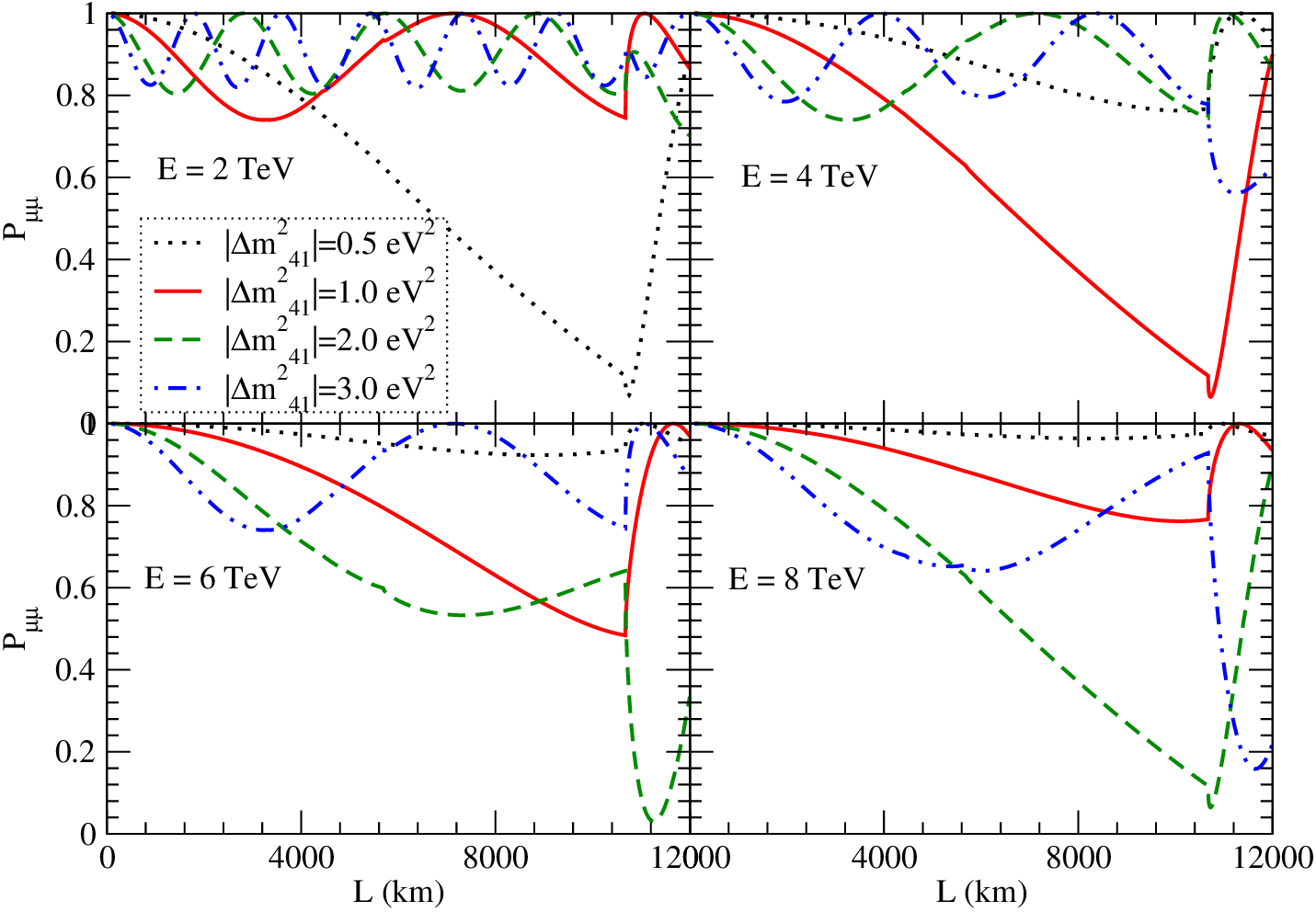}
\caption{\label{fig:Lfixedd41}
The survival probability $P_{\mu\mu}$ as a function of 
$L$ using the PREM profile for the Earth density. 
The different line types correspond to different fixed 
values of $|\mss|$ and each panel shows the results for different 
fixed values of $E$ given in the panels. 
We have assumed the 3+1 mass 
spectrum and taken $\sms=0.04$ eV$^2$, $\ses=0$ and $\sts=0$. 
The probability corresponds to neutrinos for $\mss < 0$ and 
to antineutrinos for $\mss > 0$.
}
\end{center}
\end{figure}

We start by showing results for the case where there is only 
one extra sterile neutrino. 
For 4 neutrinos we have 3 mass squared differences. 
For $\ms$ and $\ma$ we take the 
current best-fit values coming from global neutrino 
oscillation data, while for $\Delta m_{41}^2$ we take  
different values in the eV$^2$ range. 
For the mixing matrix we choose the 
following convention:
\be
U = R(\theta_{34})R(\theta_{24})R(\theta_{23})
R(\theta_{14})R(\theta_{13})R(\theta_{12})~,
\label{eq:us3p1}
\ee
where $R(\theta_{ij})$ are the rotation matrices 
and $\theta_{ij}$ the mixing angle. 
In general for the 3+1 scenario there are 3 CP violating 
Dirac phases. However
as mentioned before, 
we have put all phases to zero in Eq. (\ref{eq:us3p1}) for 
simplicity. 
The expressions for the oscillation probabilities 
relevant for atmospheric neutrinos then 
take the simple form
\be
P_{\mu\mu} &\simeq& 1 - \bigg(\sin^2\theta_{24}^M\sin^22\theta_{14}^M
+ \cos^2\theta_{14}^M\sin^22\theta_{24}^M\bigg)\,\sin^2
\bigg[\frac{(\Delta m_{41}^2)^M L}{4E}\bigg]\,,
\\
P_{\mu e} &\simeq& \sin^22\theta_{14}^M\smsm\,\sin^2
\bigg[\frac{(\Delta m_{41}^2)^M L}{4E}\bigg]\,,
\\
P_{\mu \tau} &\simeq& \cos^2\theta_{14}^M\sin^22\theta_{24}^M\stsm\,\sin^2
\bigg[\frac{(\Delta m_{41}^2)^M L}{4E}\bigg]\,,
\\
P_{\mu s} &\simeq& \cos^2\theta_{14}^M\sin^22\theta_{24}^M\cos^2\theta_{34}^M\,
\sin^2\bigg[\frac{(\Delta m_{41}^2)^M L}{4E}\bigg]\,,
\\
P_{ee} &\simeq& 1 - \sin^22\theta_{14}^M\sin^2
\bigg[\frac{(\Delta m_{41}^2)^M L}{4E}\bigg]\,,
\\
P_{e\tau } &\simeq& \sin^22\theta_{14}^M \cos^2\theta_{24}^M \stsm \,
\sin^2\bigg[\frac{(\Delta m_{41}^2)^M L}{4E}\bigg]\,.
\ee
We note that the probabilities depend only on the 
3 extra mixing angles 
$\theta_{14}$, $\theta_{24}$ and $\theta_{34}$.
In particular, $P_{\mu e}$ and $P_{\mu\mu}$ depend explictly on 
$\theta_{14}$ and $\theta_{24}$ and it seems that it is {\it apparently} 
independent of $\theta_{34}$. We will see that this is not the case 
always and there is an implicit $\theta_{34}$ dependence 
due to matter effects. 

The mixing angle $\theta_{14}$ affects strongly the 
$P_{ee}$ channel. But 
in this section, we keep fixed $\theta_{14}=0$ for simplicity and 
concentrate on the oscillation channels affecting the muon 
type (anti)neutrino. We will probe the impact of $\theta_{14}$ 
in the more realistic 3+2 scenario in next section. 
We first present results where the mixing angle 
$\theta_{34}$ is also fixed at 0 and only $\theta_{24}$ 
is the non-zero sterile mixing angle. Finally we present 
results where both $\theta_{24}$ and $\theta_{34}$ are non-zero.

\subsection{Oscillation Probabilities in 3+1 when 
$\theta_{14}=0$ and $\theta_{34}=0$}

For both $\theta_{14}=0$ and $\theta_{34}=0$, the 
probabilities assume very simple forms
\be
P_{\mu\mu} &\simeq& 1 - \sin^22\theta_{24}^M\,\sin^2
\bigg[\frac{(\Delta m_{41}^2)^M L}{4E}\bigg]\,,
\\
P_{\mu s} &\simeq& \sin^22\theta_{24}^M\,
\sin^2\bigg[\frac{(\Delta m_{41}^2)^M L}{4E}\bigg]\,,
\label{eq:ms34zero}
\\
P_{\mu e} &\simeq& 0\,,~~
P_{\mu \tau} \simeq 0\,,~~
P_{e\tau} \simeq 0\,,~~
P_{ee} \simeq 1\,.
\label{eq:mt34zero}
\ee
This is therefore a case of simple two-generation 
$\numu-\nu_s$ oscillations.
The mixing angle and mass squared difference in matter are given as
\be
\sin 2\theta_{24}^M = \sin 2\theta_{24}\frac{\mss}{(\mss)^M}\,,
\ee
\be
(\mss)^M = \sqrt{(\mss\cos2\theta_{24} \pm \sqrt{2}G_FN_A\rho Y_eE)^2
+(\mss\sin2\theta_{24})^2}\,, 
\ee
where the + sign is for neutrino and $-$ sign for the antineutrinos. 
Note that we have used $\mss \simeq \Delta m^2_{42}$.   
As discussed before, we have resonant matter effects 
and $\sin^22\theta_{24}^M=1$ 
in the neutrino (antineutrino) channel when $\mss < 0$ ($\mss >0$).
However, the condition of resonance does not necessarily give the 
largest possible oscillations. 
The condition for maximal oscillation is achieved 
when both $\sin^22\theta_{24}^M=1$ and 
$\sin^2[(\Delta m_{41}^2)^M L/4E]=1$ simultaneously 
\cite{maxmatter} and is given by Eq. (\ref{eq:maxoscgen}). 
It 
can be inverted to give the value of $\theta_{24}$ which 
would give maximal oscillations at a given baseline:
\be
\tan2\theta_{24} = \frac{32.55\times 10^3}{\rho L~({\rm km~gm/cc})}~.
\label{eq:max24}
\ee
For values of $\theta_{24}$ either less or greater than the 
value corresponding to
that obtained from Eq. (\ref{eq:max24}), 
the oscillations are less \cite{bino}. 
We show in Fig. \ref{fig:maxosc} the value of 
$\sms$ for which we can have 
maximal oscillations, as a function of distance $L$ travelled 
inside Earth. For $\rho$ we have used the average 
density along the neutrino trajectory given by the PREM 
profile. The dashed line shows the mantle-core boundary of 
Earth and we can see that for the more plausible values of 
$\sms\ltap 0.07$ the condition for maximal oscillations are always 
met inside the Earth's core. In particular, for the longest 
possible trajectory where $L=2\times R_E$, $R_E$ being the 
Earth's radius, we note that maximal oscillations will happen 
if $\sms=0.02$. 

Fig. \ref{fig:LfixedE} shows the survival probability 
$P_{\mu\mu}$ as a function of the distance travelled inside the
Earth. For this plot we use the full PREM density profile 
for the Earth and solve the four neutrino differential equation 
in matter. 
The different line types correspond to different fixed 
values of $E$ and each panel shows the results for different 
fixed values of $\sms$ given in the panels. 
The probability 
shown would correspond to that for neutrinos if $\mss < 0$ and 
to antineutrinos if $\mss > 0$. 
We note from the figure that for most neutrino energies 
above $E\gtap 2$ TeV, there are sizeable matter effects inside 
Earth and 
the survival probability generally decreases with $L$ in 
the mantle. Inside the core it falls first, followed by a rise. 
However, we can see that for reasonable $L$ binning of the 
high energy atmospheric neutrino data in neutrino 
telescopes, it should be possible to see 
zenith angle dependent fall in 
$P_{\mu\mu}$. Even for very small values of $\sms$ like 0.01, we 
can see that $P_{\mu\mu}$ could fall to up to 0.6, and this should be 
observable in the detector. 

Fig. \ref{fig:Lfixedd41} is similar to Fig. \ref{fig:LfixedE}, 
except that here we show the probability at a fixed 
value of $\sms$, but different choices of $|\mss|$ and 
$E$. All plots are for $\sms=0.04$ and each panel for 
a fixed $E$, shown in the figure. The different line 
types correspond to different $|\mss|$. As before, 
the probability corresponds to neutrinos for $\mss < 0$ and 
to antineutrinos for $\mss > 0$. This figure tells us how 
different values of $|\mss|$ can be distinguished from the 
high energy atmospheric neutrino data at the neutrino 
telescopes. We can see that binning in 
either or both $E$ and $L$ would 
help in distinguishing between the different possible 
$|\mss|$ values. 

\subsection{Oscillation Probabilities in 3+1 when 
$\theta_{14}=0$ and $\theta_{34}\neq0$}

\begin{figure}[t]
\begin{center}
\includegraphics[width=16.0cm, height=9.0cm, angle=0]{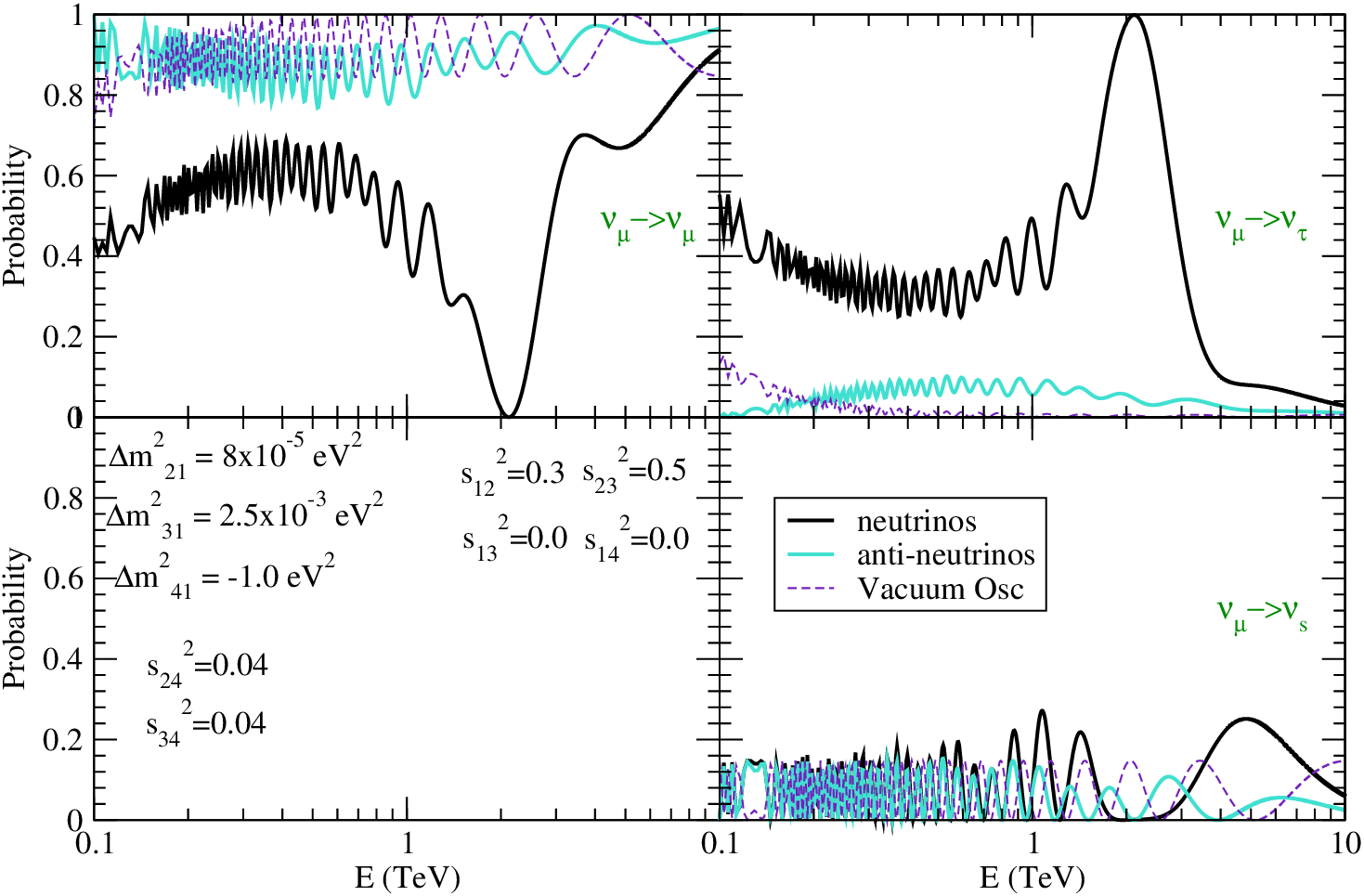}
\caption{\label{fig:EIH}
The $\numu\rightarrow \numu$ (upper left hand panel), 
$\numu\rightarrow \nutau$ (upper right hand panel)
and $\numu\rightarrow \nu_s$ (lower right hand panel)
oscillation probabilities, as a function of the 
neutrino energy $E$ for the 3+1 mass spectrum, when the 
neutrinos travel a distance $L=2R_e$, where $R_e$ is the 
radius of the Earth.  
The black (dark) solid lines show the 
probabilities for neutrinos while the cyan (light) solid 
lines show the probabilities for antineutrinos. The 
dashed lines give the probabilities if matter effects were 
not taken into account and one had oscillations in vacuum.
The values of the oscillation parameters taken for this figure 
is shown in the  lower left hand panel. In particular this plot 
is for $\mss =1$ eV$^2$. Probabilities in matter have been obtained 
using the PREM profile.
}
\end{center}
\end{figure}

\begin{figure}[t]
\begin{center}
\includegraphics[width=16.0cm, height=9.0cm, angle=0]{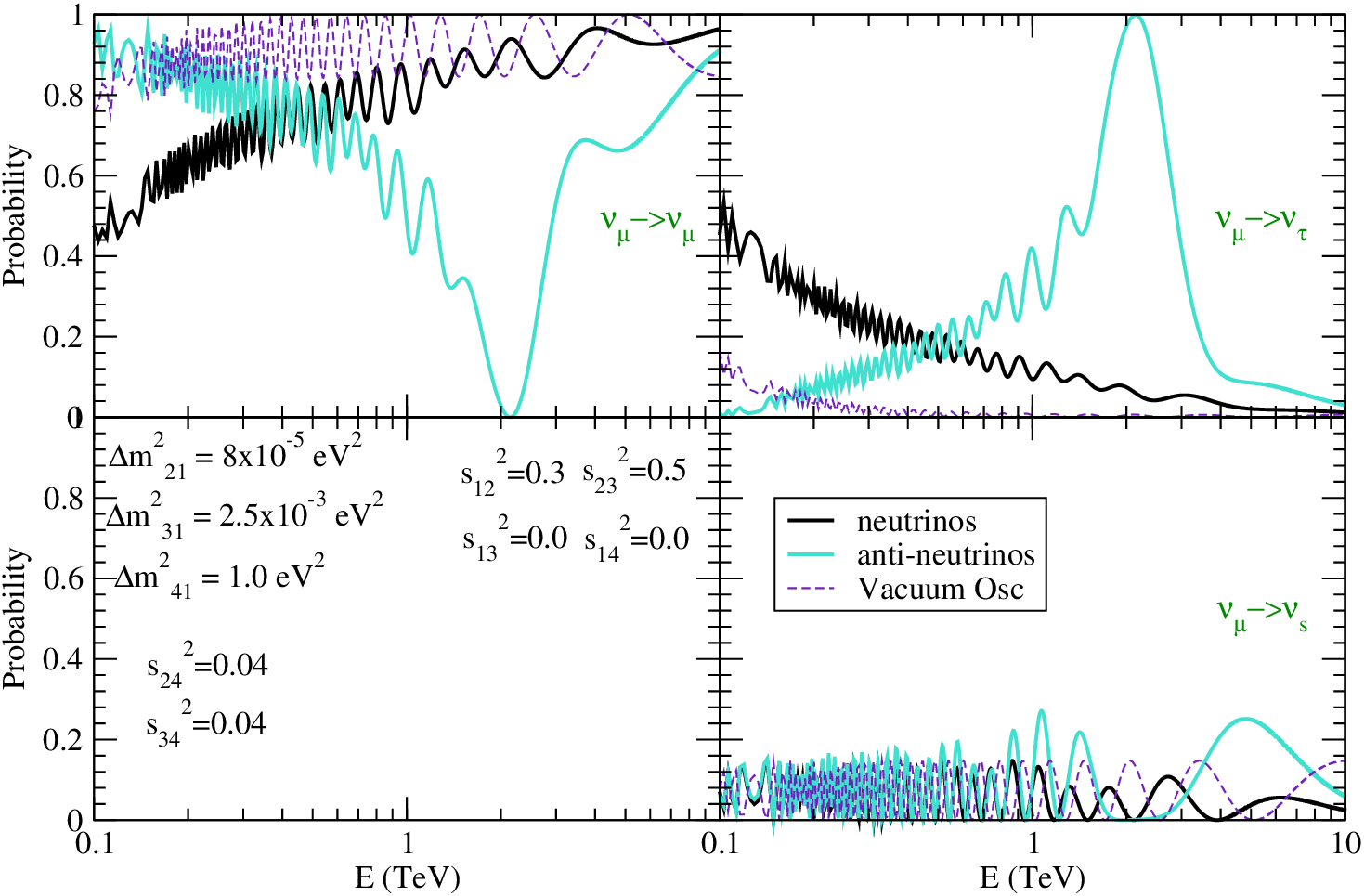}
\caption{\label{fig:ENH}
Same as Fig. \ref{fig:EIH} but for $\mss = +1$ eV$^2$. 
}
\end{center}
\end{figure}

\begin{figure}[t]
\begin{center}
\includegraphics[width=16.0cm, height=9.0cm, angle=0]{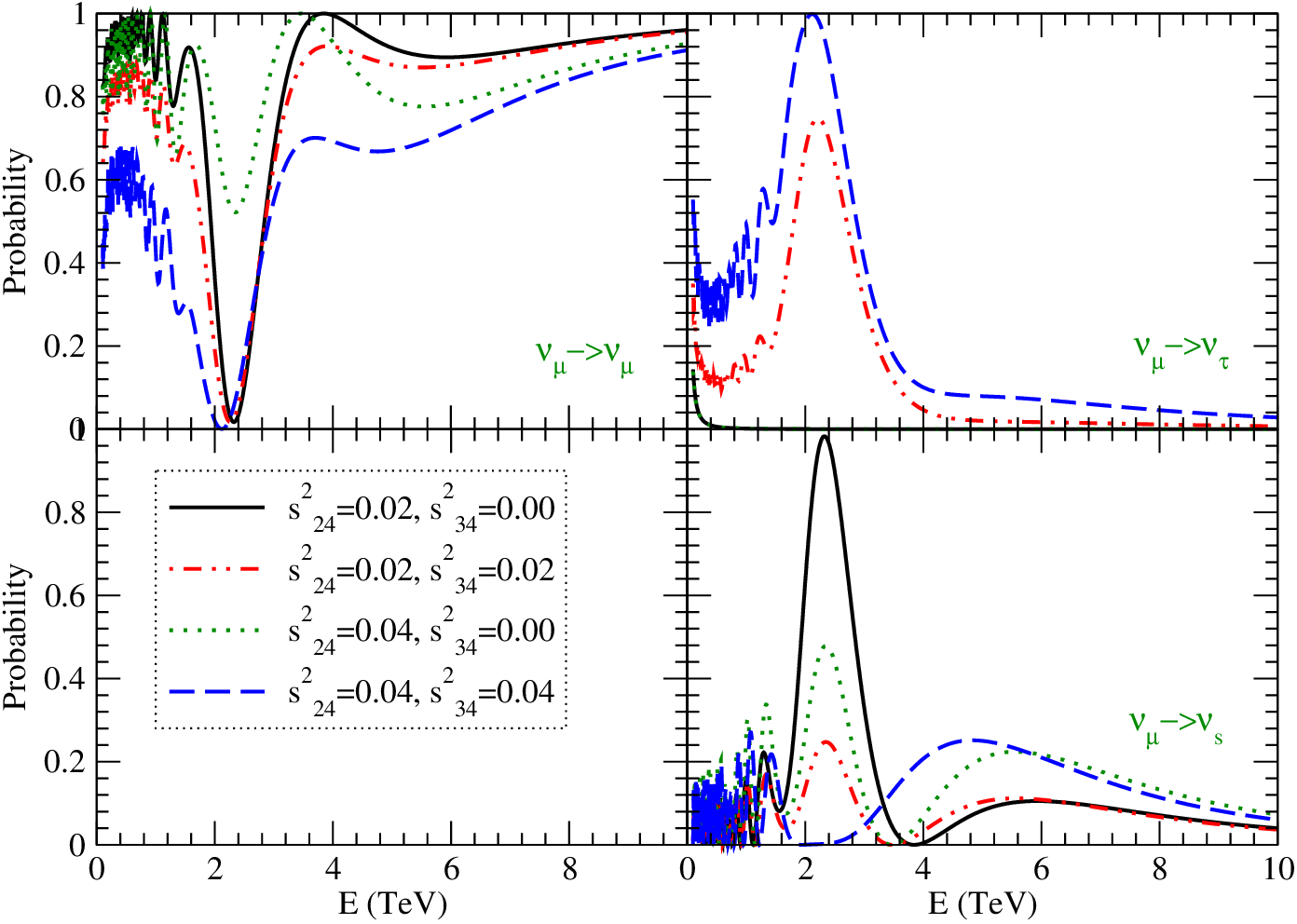}
\caption{\label{fig:EIH_24_34}
The $\numu\rightarrow \numu$ (upper left hand panel), 
$\numu\rightarrow \nutau$ (upper right hand panel) 
and $\numu\rightarrow \nu_s$ (lower right hand panel)
oscillation probabilities, as a function of the 
neutrino energy $E$, when the 
neutrinos travel a distance $L=2R_e$, where $R_e$ is the 
radius of the Earth. Different line types correspond to 
different combinations of $\sms$ and $\sts$ and 
$\mss = -1$ eV$^2$.  
}
\end{center}
\end{figure}

\begin{figure}[t]
\begin{center}
\includegraphics[width=10.0cm, height=7.0cm, angle=0]{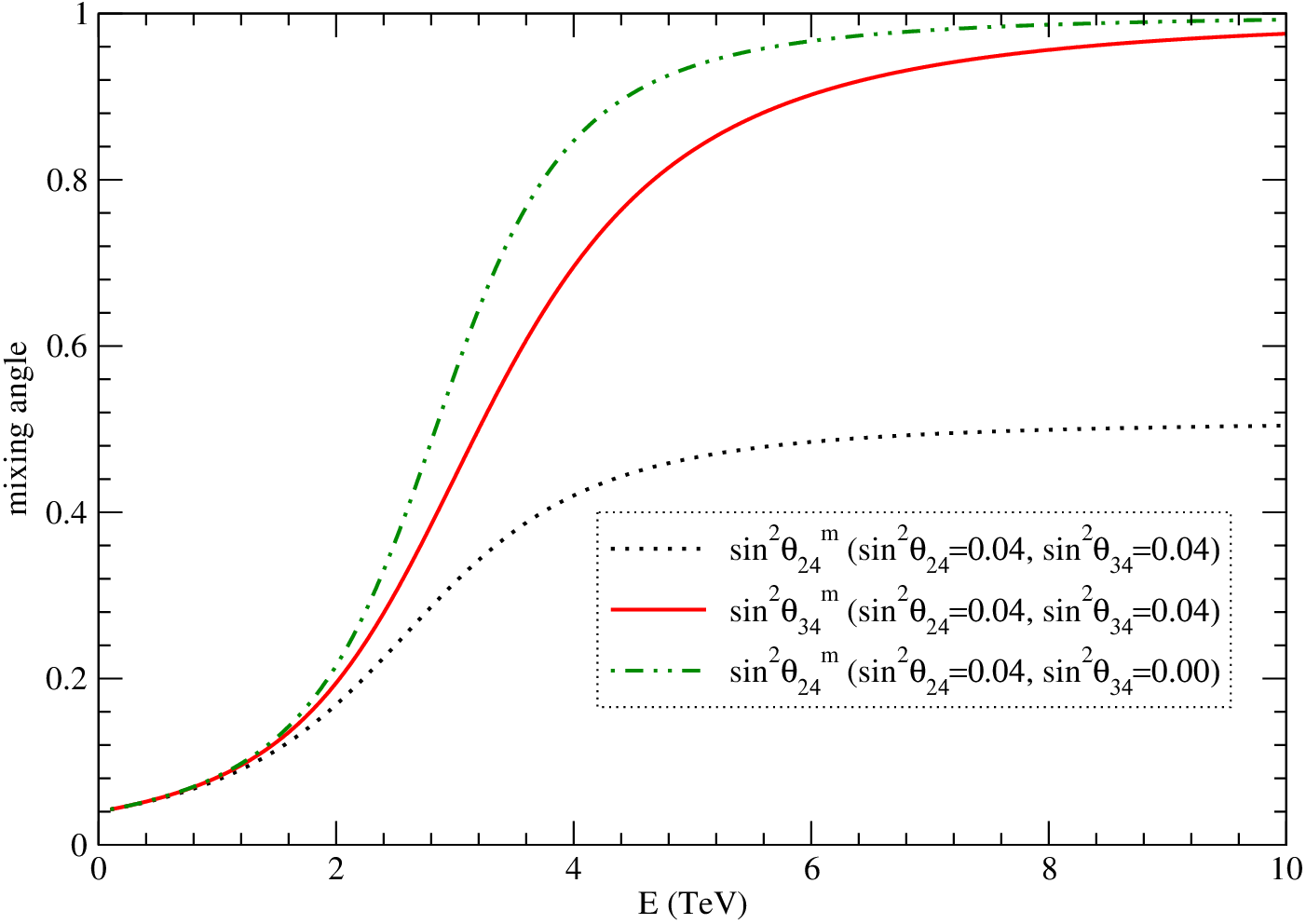}
\caption{\label{fig:mixingangle}
The mixing angles in matter as a function of the neutrino energy. 
The dotted and solid lines show $\smsm$ and $\stsm$ 
when their values in vacuum are $\sms=0.04$ and $\sts=0.04$.
The dotted-dashed lines show $\smsm$ 
when the values in vacuum are $\sms=0.04$ and $\sts=0.00$.
}
\end{center}
\end{figure}

If we allow $\sts \neq 0$ but still keep $\theta_{14}=0$, 
the probabilities are given as
\be
P_{\mu\mu} &\simeq& 1 - \sin^22\theta_{24}^M\,\sin^2
\bigg[\frac{(\Delta m_{41}^2)^M L}{4E}\bigg]\,,
\\
P_{\mu \tau} &\simeq& \sin^22\theta_{24}^M\stsm\,\sin^2
\bigg[\frac{(\Delta m_{41}^2)^M L}{4E}\bigg]\,,
\\
P_{\mu s} &\simeq& \sin^22\theta_{24}^M\cos^2\theta_{34}^M\,
\sin^2\bigg[\frac{(\Delta m_{41}^2)^M L}{4E}\bigg]\,,
\\
P_{\mu e} &\simeq& 0\,,~~
P_{e \tau} \simeq 0\,,~~
P_{ee} \simeq 1\,.
\ee
Fig. \ref{fig:EIH} shows the neutrino oscillation probabilities 
assuming inverted mass ordering, taking 
$\mss=-1$ eV$^2$. The $\numu\rightarrow \numu$, 
$\numu\rightarrow \nutau$ and $\numu\rightarrow \nu_s$ 
probabilities are shown in the 
upper left hand panel, upper right hand panel and 
lower right hand panel, respectively. For $\ses=0$, 
the probabilities  $P_{\mu e} \simeq 0$, $P_{e \tau} \simeq 0$ and 
$P_{ee} \simeq 1$ and hence we do not show them.
The solid black (dark) line is for neutrinos in matter and 
the solid cyan (light) line is for antineutrinos in matter, 
while the thin black dashed line shows the probabilities in 
vacuum for comparison. 
We stress that even though we have denoted the probabilities 
as $\numu\rightarrow \numu$ etc. in the figure, 
its understood that we are 
using them to denote the probability for both the neutrino 
as well as the antineutrino channels.
We have kept $\sms=\sts=0.04$ in this figure. The corresponding 
plots with the normal mass ordering is 
shown in Fig. \ref{fig:ENH}. The neutrinos (antineutrinos) 
undergo maximal oscillations around $E=2$ TeV when 
$\mss =-1$ eV$^2$ ($\mss=+1$ eV$^2$). 
At lower values of $E$, where very large matter effects 
in $\mss$ oscillations have still not 
set in, we note a marked difference between the oscillations 
pattern of neutrinos and antineutrinos and between the cases where 
$\mss <0$ and $\mss >0$. The oscillations for lower $E$ are 
dependent on $\ma$ as well 
and the difference between the oscillation 
mentioned above is due to 
both $\ma$ and $\mss$ dependent terms. Note that 
in both cases we have kept $\ma >0$. 
The most important thing to note from this figure is that 
around the point where we have 
maximal matter effects, $P_{\mu\mu}\simeq 0$, $P_{\mu\tau}\simeq 1$ and 
$P_{\mu s}\simeq 0 $ for the neutrino (antineutrino) channel 
for $\mss <0$ ($\mss >0$). Such large oscillations should not 
be difficult to observe in the very high atmospheric neutrino
data in neutrino telescopes.

In the previous subsection where we had put $\theta_{34}=0$, 
we had argued 
that for $L=2R_e$, where $R_e$ is the Earth's radius, 
maximal oscillations of $\numu$ would occur around $\sms \simeq 0.02$. 
For $\sms=0.04$ we should therefore expect lesser oscillations.
We had drawn these conclusions using average 
constant matter density approximation for the Earth.
For the PREM profile which 
corresponds to varying matter density for the Earth, this 
scenario holds, albeit approximately. In Figs. \ref{fig:EIH} and 
\ref{fig:ENH} we can 
see that we get maximal oscillations even for 
$\sms=0.04$ as long as $\sts=0.04$. 
Another important aspect 
we note from the figures is 
that $P_{\mu\tau} \simeq 1$ and $P_{\mu s}\simeq 0$ when 
we have maximal oscillations of $\numu$. 
This is in stark contrast to the case where $\sts=0$, for 
which we had $P_{\mu\tau}\simeq 0$ and $P_{\mu s}\simeq 1$
(cf. Eqs. (\ref{eq:ms34zero}) and (\ref{eq:mt34zero})). 
The main reason 
for this complete reversal of scenario is that when $\sms\neq 0$ 
and $\sts=0$, $\smsm$ is enhanced in matter for neutrinos 
(antineutrinos) when $\mss <0$ ($\mss > 0$) while 
$\stsm$ remains zero. Therefore, $P_{\mu\tau}$ is 
always zero and 
we have simple two-generation matter enhanced 
$\numu\rightarrow \nu_s$ oscillations. However, 
when $\sts \neq 0$, both $\sms$ and $\sts$ are enhanced 
in matter for neutrinos 
(antineutrinos) when $\mss <0$ ($\mss > 0$). 
This is a genuine three-generation oscillation case in 
which if $\sts=\sms$,  
the $\numu$ and $\nutau$ states 
evolve identically in matter and 
resonate with the sterile state at almost the same 
energy. 

To further illustrate this point we present Fig. \ref{fig:EIH_24_34},
where we compare the probabilities corresponding to $L=2R_e$,
for the cases where $\sts=0$ 
with those where $\sts\neq 0$. We show $P_{\mu\mu}$, 
$P_{\mu\tau}$ and $P_{\mu s}$ for $\sms=0.02$, $\sts=0.00$
(black solid lines), 
$\sms=0.02$, $\sts=0.02$
(red dot-dashed lines),
$\sms=0.04$, $\sts=0.00$
(green dotted lines), and 
$\sms=0.04$, $\sts=0.02$
(blue dashed lines). 
We reconfirm that for $\sms=0.02$ and $\sts=0$ we have 
$P_{\mu\mu}\simeq 0$ and 
$P_{\mu s}\simeq 1$. 
For $\sms=0.04$ and $\sts=0$, we still have two-generation 
$\numu \rightarrow \nu_s$ oscillations with $P_{\mu\tau}=0$, 
but now since we have shifted from the most optimal $\sms$ value 
for this baseline, $P_{\mu\mu}$ increases and $P_{\mu s}$ 
decreases compared to the case where 
$\sms=0.02$. Once we put $\sts=\sms$, we get 
$P_{\mu\mu}\simeq 0$ for both $\sms=0.02$ and $\sms=0.04$. 
However, non-zero $\sts$ brings a huge change in 
$P_{\mu\tau}$ which becomes
non-zero and large 
and for  $\sts=0.04$, it is in fact very close to 1. 
Likewise, $P_{\mu s}$ changes substantially due to $\sts$. 

In order to quantify our discussion on the impact of $\sts$
on the evolution of the 
neutrino states inside Earth, 
we show in Fig. \ref{fig:mixingangle}
the mixing angles in matter as a function of $E$. Since its not 
possible to show the evolution of the mixing angles for the 
full PREM profile, we show a snapshot for a density of 
$\rho=8.44$ gm/cc, which is the average density encountered by a 
neutrino moving along the diameter of the Earth. The green 
dot-dashed line shows $\sin^2\theta_{24}^M$ when 
$\sms=0.04$ and $\sts=0$ in vacuum. We see that the mixing 
angle increases with energy and hence matter effects. At 
the resonance energy we get $\sin^22\theta_{24}^M=1$ and beyond
that $\sin^2\theta_{24}^M$ keeps increasing to 1 and 
$\sin^22\theta_{24}^M$ decreases. The evolution of 
$\sin^2\theta_{24}^M$ when 
$\sms=0.04$ and $\sts=0.04$ is shown by the black dotted line. 
This case is very different from the earlier described case. 
Here $\sin^2\theta_{24}^M$ remains more or less constant beyond the 
resonance and hence $\sin^22\theta_{24}^M$ assumes some 
large constant value and does not decrease like before. 
Note that the reason we were getting lesser oscillations in the 
$P_{\mu\mu}$ channel for $\sms=0.04$ (and $\sts=0$) was because 
the resonance energy here was not matching exactly with the 
energy as which the oscillatory term peaks. But for 
$\sms=0.04$ and $\sts=0.04$ since 
$\sin^22\theta_{24}^M$ has a large value from energies beyond the 
resonance, the problem of fine tuning the resonance energy and 
oscillations peak energy is drastically reduced and we can have 
$P_{\mu\mu} \simeq 0$ more easily. 
The reason for the change in behavior of  
$P_{\mu\tau}$ and $P_{\mu s}$ with $\sts \neq 0$ 
can also be seen from 
Fig. \ref{fig:mixingangle}. 
The red solid line shows the angle $\sin^2\theta_{34}^M$ and 
we see that this keeps increasing and goes to 1 for 
$E$ greater than the resonance energy.
Since $P_{\mu\tau}$ is 
proportional to $\sin^2\theta_{34}^M$ and 
$P_{\mu s}$ to $\cos^2\theta_{34}^M$, 
$P_{\mu\tau}$ increases while $P_{\mu s}$ decreases as 
$\sin^2\theta_{34}^M$ increases.

\section{Neutrino Oscillations with 3+2 Mass Spectrum}

\begin{figure}[t]
\begin{center}
\includegraphics[width=16.0cm, height=8.8cm, angle=0]{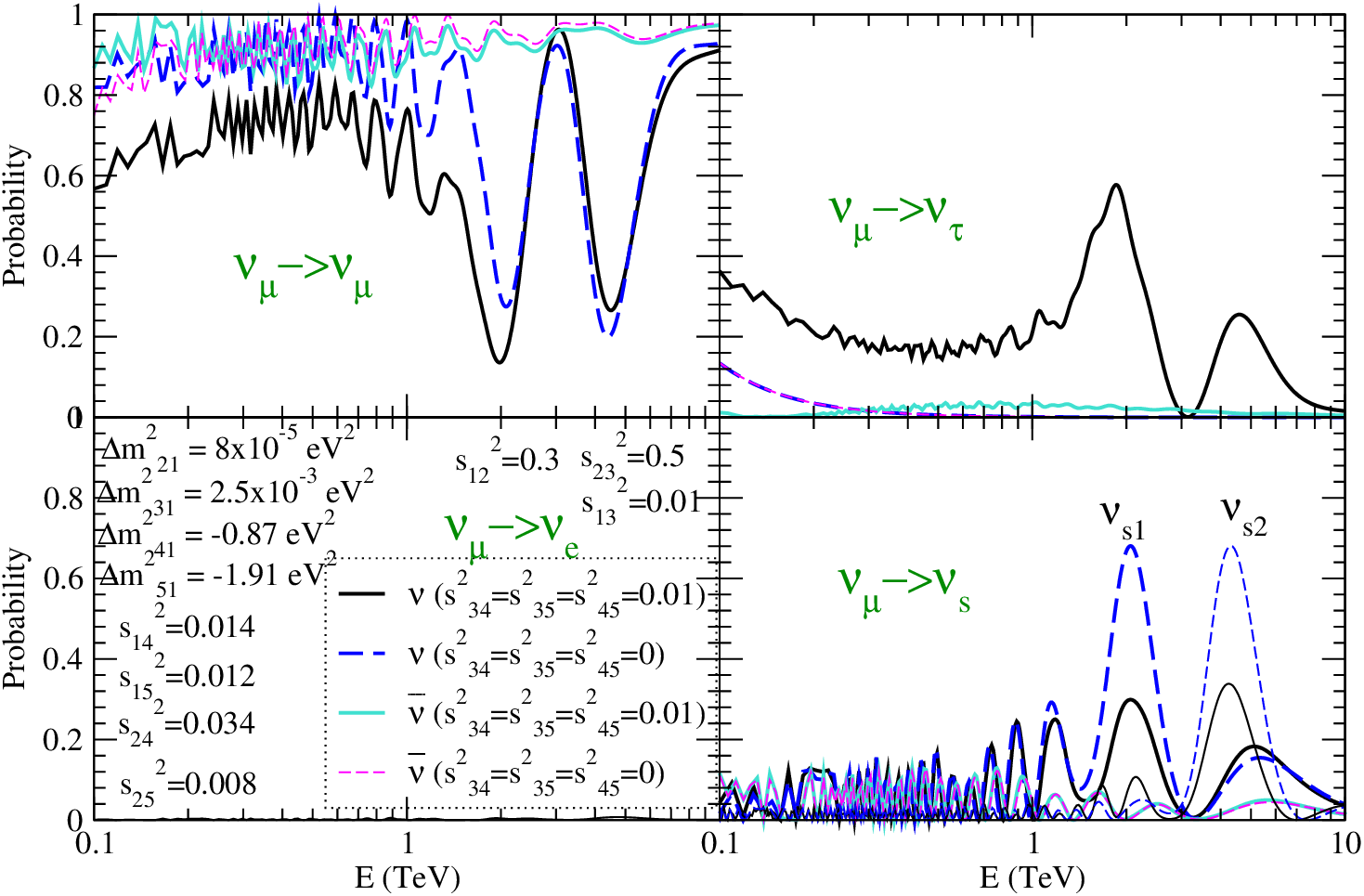}
\caption{\label{fig:3p2EIH}
The $\numu\rightarrow \numu$ (upper left hand panel), 
$\numu\rightarrow \nutau$ (upper right hand panel), 
$\numu\rightarrow \nue$ (lower left hand panel), and
$\nue\rightarrow \nu_{s1}$ and
$\nue\rightarrow \nu_{s2}$  (lower right hand panel)
oscillation probabilities, as a function of the 
neutrino energy $E$ for the 3+2 mass spectrum, when the 
neutrinos travel a distance $L=2R_e$, where $R_e$ is the 
radius of the Earth. 
The solid black and dashed blue lines show the 
probabilities for neutrinos while 
the solid cyan and thin dashed magenta lines are for the 
antineutrinos. The solid black and solid cyan lines are drawn 
for $\sts=\stss=\sin^2\theta_{45}=0.01$, while 
the dashed blue and dashed magenta lines are for 
$\sts=\stss=\sin^2\theta_{45}=0.0$.  
The values of the other 
oscillation parameters are shown in the figure and we have taken 
$\mss=-0.87$ eV$^2$ and 
$\msss=-1.91$ eV$^2$. For $\numu\rightarrow \nu_s$, we show the 
$\numu\rightarrow \nu_{s1}$ by thick line and 
$\numu\rightarrow \nu_{s2}$ by thin line.
}
\end{center}
\end{figure}

\begin{figure}[t]
\begin{center}
\includegraphics[width=16.0cm, height=10.0cm, angle=0]{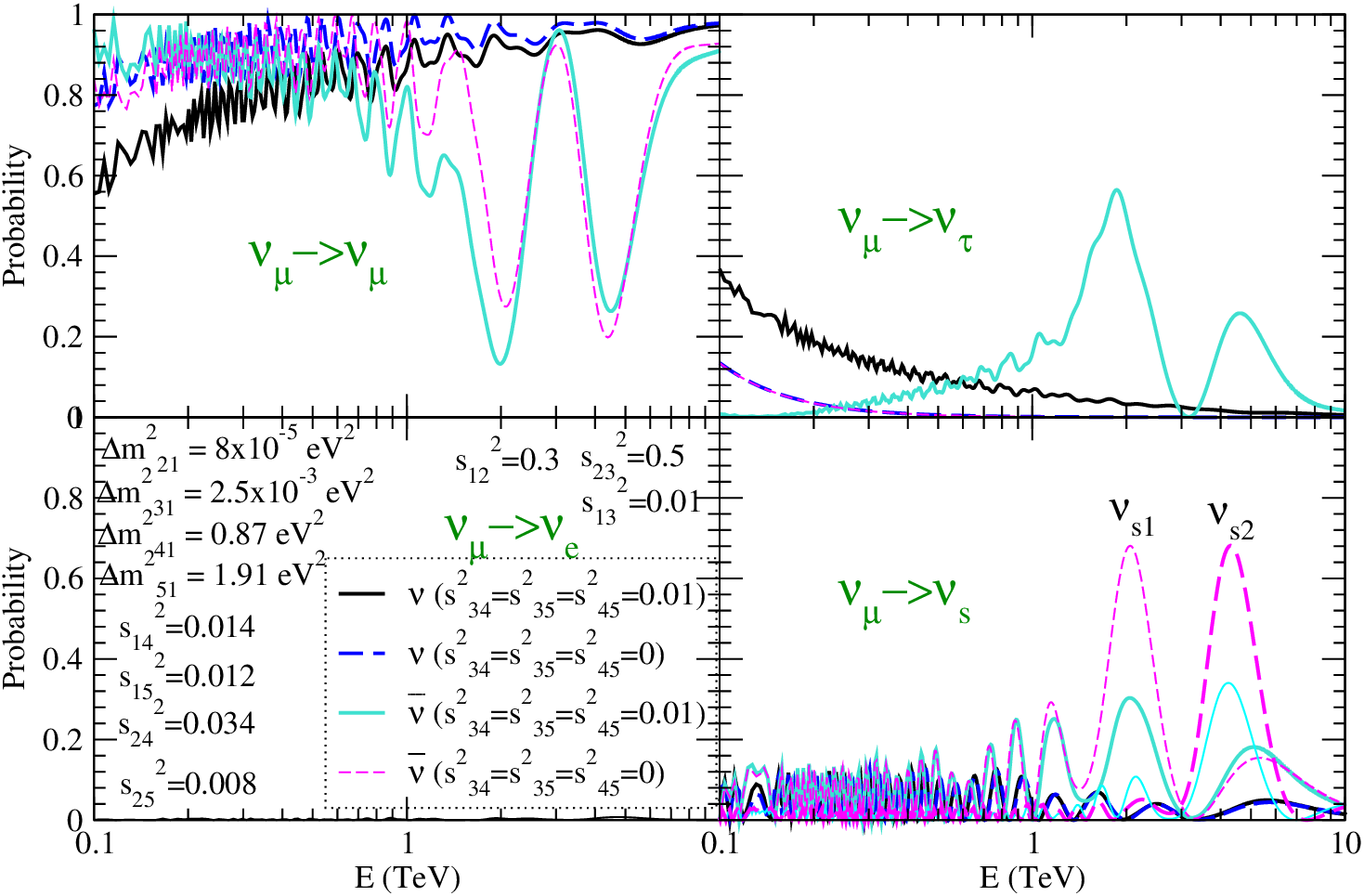}
\caption{\label{fig:3p2ENH}
Same as Fig. \ref{fig:3p2EIH} but for $\mss=+0.87$ eV$^2$ and 
$\msss=+1.91$ eV$^2$. For $\numu\rightarrow \nu_s$, we show the 
$\numu\rightarrow \nu_{s1}$ by thin line and 
$\numu\rightarrow \nu_{s2}$ by thick line.
}
\end{center}
\end{figure}

\begin{figure}[t]
\begin{center}
\includegraphics[width=16.0cm, height=9.0cm, angle=0]{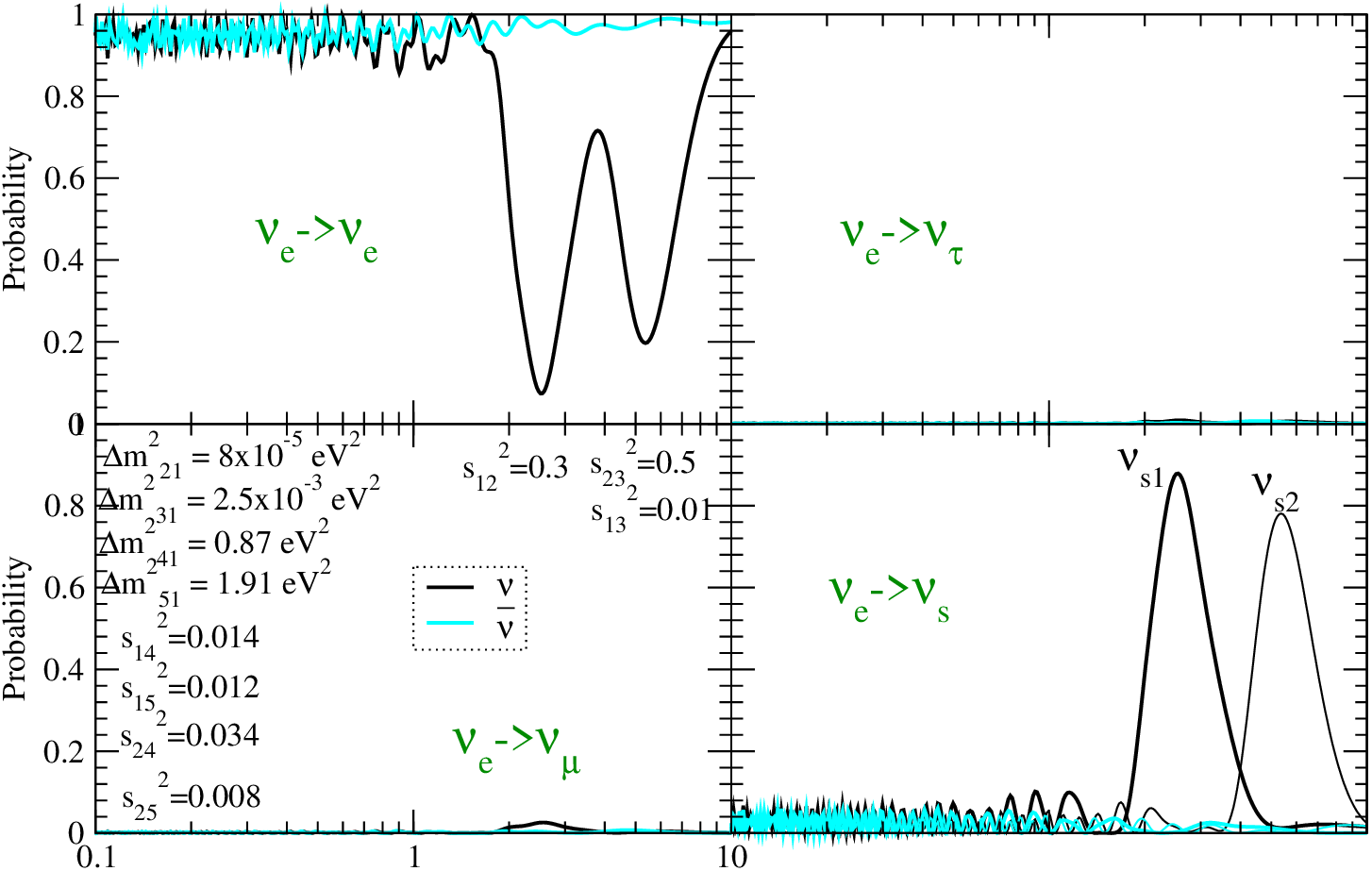}
\caption{\label{fig:3p2Eee}
The $\nue\rightarrow \nue$ (upper left hand panel), 
$\nue\rightarrow \nutau$ (upper right hand panel), 
$\nue\rightarrow \numu$ (lower left hand panel), and
$\nue\rightarrow \nu_{s1}$ and
$\nue\rightarrow \nu_{s2}$  (lower right hand panel)
oscillation probabilities, as a function of the 
neutrino energy $E$ for the 3+2 mass spectrum, when the 
neutrinos travel a distance $L=2R_e$, where $R_e$ is the 
radius of the Earth. 
The solid black lines show the 
probabilities for neutrinos while 
the solid cyan lines are for the 
antineutrinos. For $\nue\rightarrow \nu_s$, we show the 
$\nue\rightarrow \nu_{s1}$ by thick line and 
$\nue\rightarrow \nu_{s2}$ by thin line.
We have taken 
$\mss=0.87$ eV$^2$ and 
$\msss=1.91$ eV$^2$. The oscillation probabilities mainly depend on
only $\ses$ and $\sess$ and almost independent of all other 
mixing angles.   
}
\end{center}
\end{figure}

The 3+1 neutrino mass spectrum, though simpler for the 
understanding of the resonant oscillation picture, stands disfavored
comprehensively once the latest MiniBooNE results are 
included into the analysis along with results from all 
other neutrino oscillation experiments. However, the 3+2 scheme, with 
two extra sterile neutrinos, provides a very reasonable 
description of the world neutrino data, including MiniBooNE. 
In this section we look at the predicted neutrino and antineutrino 
oscillation probabilities at neutrino telescopes for the 
3+2 neutrino mass scheme. The expression for the oscillation 
probabilities in the 3+2 scheme are as follows:
\be
P_{\alpha\alpha} = 1 &-& 4|U_{\alpha 4}^M|^2\bigg(1-|U_{\alpha 4}^M|^2\bigg)
\sin^2\bigg(\frac{(\mss)^M L}{4E}\bigg)
- 4|U_{\alpha 5}^M|^2\bigg(1-|U_{\alpha 5}^M|^2\bigg)
\sin^2\bigg(\frac{(\msss)^M L}{4E}\bigg)
\nonumber\\
&+&8|U_{\alpha 4}^M|^2|U_{\alpha 5}^M|^2
\sin\bigg(\frac{(\mss)^M L}{4E}\bigg)
\sin\bigg(\frac{(\msss)^M L}{4E}\bigg)
\sin\bigg(\frac{(\Delta m^2_{54})^M L}{4E}\bigg)\,,
\label{eq:surv3p2}
\ee
\be
P_{\alpha\beta} = &&4|U_{\alpha 4}^M|^2|U_{\beta 4}^M|^2
\sin^2\bigg(\frac{(\mss)^M L}{4E}\bigg)
+ 4|U_{\alpha 5}^M|^2|U_{\beta 5}^M|^2
\sin^2\bigg(\frac{(\msss)^M L}{4E}\bigg)\,.
\nonumber\\
&+&
 8 |U_{\alpha 4}^M{U_{\beta 4}^M}^*{U_{\alpha 5}^M}^*U_{\beta 5}^M|
\sin\bigg(\frac{(\mss)^M L}{4E}\bigg)
\sin\bigg(\frac{(\msss)^M L}{4E}\bigg)
\cos\bigg(\frac{(\Delta m^2_{54})^M L}{4E}\bigg)
\label{eq:tran3p2}
\ee
Note that if we had taken into account CP violating phases in 
$U$, the argument in the cosine of the last term in Eq. 
(\ref{eq:tran3p2}) would be $(\Delta m^2_{54})^M L/4E - \delta_{\alpha\beta}$
where $\delta_{\alpha\beta} = Arg(
U_{\alpha 4}^M{U_{\beta 4}^M}^*{U_{\alpha 5}^M}^*U_{\beta 5}^M)$.

For the three active and 
two sterile neutrino framework we 
have 4 independent mass squared differences and 
hence can have the following 
possibilities for the mass spectrum which we call \cite{sn3p2}:
\be
{\rm N2+N3:}~ &\ma>0,& \Delta m^2_{41}>0 ~{\rm and}~ \Delta m^2_{51}>0~, \label{eq:n2n3}\\
{\rm N2+I3:}~ &\ma<0,& \Delta m^2_{41}>0 ~{\rm and}~ \Delta m^2_{51}>0~, \label{eq:n2i3}\\
{\rm H2+N3(a):}~ &\ma>0,& \Delta m^2_{41}>0 ~{\rm and}~ \Delta m^2_{51}<0~, \label{eq:h2n3}\\
{\rm H2+I3(a):}~ &\ma<0,& \Delta m^2_{41}>0 ~{\rm and}~ \Delta m^2_{51}<0~, \label{eq:h2i3}\\
{\rm I2+N3:}~ &\ma>0,& \Delta m^2_{41}<0 ~{\rm and}~ \Delta m^2_{51}<0~, \label{eq:t2n3}\\
{\rm I2+I3:}~ &\ma<0,& \Delta m^2_{41}<0 ~{\rm and}~ \Delta m^2_{51}<0~,\label{eq:i2i3}
\ee
with $\ms>0$ always. In addition, 
the H2+N3 and H2+I3 schemes can have 
2 more possibilities \cite{ws3p2}
\be
{\rm H2+N3(b):}~ &\ma>0,& \Delta m^2_{41}<0 ~{\rm and}~ \Delta m^2_{51}>0~, \label{eq:h2n3n}\\
{\rm H2+I3(b):}~ &\ma<0,& \Delta m^2_{41}<0 ~{\rm and}~ \Delta m^2_{51}>0~. \label{eq:h2i3n}
\ee
Since there are two mass squared difference associated with the 
sterile states, we expect two resonances. Whether 
the resonance occurs in the neutrino or the antineutrino 
channel depends on the mass ordering. 
While the ordering of the mass states within the 
three active part is almost inconsequential for the very high energy 
neutrinos we are concerned with here, the mass ordering of the 
sterile states between themselves and with respect to the 
three active states is of utmost importance. 
In particular, if both $\Delta m^2_{41}>0$ and $\Delta m^2_{51}>0$
(corresponding to the N2+N3 and N2+I3 spectra), 
then both the $\numu \rightarrow \nu_s$ ($\nue \rightarrow \nu_s$)
resonances happen in the antineutrino (neutrino) channel. 
On the other hand, if both $\Delta m^2_{41}<0$ and $\Delta m^2_{51}<0$
(corresponding to the I2+N3 and I2+I3 spectra), 
then both the $\numu \rightarrow \nu_s$ ($\nue \rightarrow \nu_s$)
resonances happen in the neutrino (antineutrino) channel.
For the hybrid cases, where 
$\Delta m^2_{41}>0$ and $\Delta m^2_{51}<0$ (H2+N3(a) and H2+I3(a))
or
$\Delta m^2_{41}<0$ and $\Delta m^2_{51}>0$ (H2+N3(b) and H2+I3(b)),
one of the resonances occur in the neutrino and another in the 
antineutrino channel.

Figs. \ref{fig:3p2EIH} and \ref{fig:3p2ENH} show the 
probabilities involving the muon (anti)neutrino, 
as a function of energy for the I2+N3 
and N2+N3 cases respectively. The solid black (dark) and dashed blue (dark)
lines are for neutrinos and solid cyan (light) 
and dashed magenta (light) lines  
are for antineutrinos. We show results for the 
global best-fit parameter values taken from \cite{maltonischwetz}.
When the entire MiniBooNE data is included in the analysis,
the authors of \cite{maltonischwetz} 
get as their best-fit $|\mss|=0.87$ eV$^2$ and 
$|\msss|=1.91$ eV$^2$ for the mass squared difference and 
$|U_{e4}|=0.12$, $|U_{e5}|=0.11$, $|U_{\mu 4}|=0.18$ and 
$|U_{\mu 5|}=0.089$. 
If we assume a parameterization for the $5\times 5$ 
mixing matrix as
\be
U = R(\theta_{45})R(\theta_{35})R(\theta_{34})R(\theta_{25})
R(\theta_{24})R(\theta_{15})
R(\theta_{14})R(\theta_{23})R(\theta_{13})R(\theta_{12})~,
\label{eq:u3p2}
\ee
then the best-fit 
values for the matrix elements mentioned above can 
be obtained if we take $\sin^2\theta_{14}=0.014$,
$\sin^2\theta_{15}=0.012$, $\sin^2\theta_{24}=0.034$ and 
$\sin^2\theta_{25}=0.008$. We present our results assuming 
these values. 
The other mixing angles associated 
with the sterile states are $\sin^2\theta_{34}$, 
$\sin^2\theta_{35}$ and $\sin^2\theta_{45}$. These remain 
almost unconstrained by the current neutrino oscillation data
and could in principle take any value. 
For the sake of illustration, 
we show results only for two sets of choices for these  
mixing angles. The solid lines show probabilities for 
$\sin^2\theta_{34}=\sin^2\theta_{35}=\sin^2\theta_{45}=0.01$, 
while the dashed lines are for 
$\sin^2\theta_{34}=\sin^2\theta_{35}=\sin^2\theta_{45}=0.0$. 
The other mixing angles are fixed at $\sin^2\theta_{12}=0.3$,
$\sin^2\theta_{23}=0.5$ and 
$\sin^2\theta_{13}=0.01$.  We see that for the I2+N3 spectrum 
both resonances are in the neutrino channel while for the 
N2+I3 spectrum both resonances are in the antineutrino channel.
We see that the effect of the mixing angles 
$\sin^2\theta_{34}$ and 
$\sin^2\theta_{35}$ is to  
increase the $\numu \rightarrow \nutau $ oscillations and 
reduce the $\numu \rightarrow \nu_s $ transitions in the resonant 
channel. The net result of these mixing angles is to reduce slightly 
the net $\numu$ survival probability. 
These features are similar to what we had observed for non-zero 
$\sin^2\theta_{34}$ for the 3+1 case discussed in the 
previous section.  The reason why $P_{\mu e}$ is very small
is easy to see from Eq. (\ref{eq:tran3p2}). 
In the neutrino channel, say, with $\mss <0$ and $\msss <0$
as in Fig. \ref{fig:3p2EIH}, while $U_{\mu 4}^M$ and 
$U_{\mu 5}^M$ increase due to resonance, $U_{e4}^M$ and 
$U_{e5}^M$ remain negligible since $\theta_{14}^M$ 
remains small, and as a result 
$P_{\mu e}$ remains negligible. In the antineutrino 
channel for this mass spectrum 
$U_{e4}^M$ and 
$U_{e5}^M$ are large (as discussed below), however in that 
case $U_{\mu 4}^M$ and 
$U_{\mu 5}^M$ are small. Therefore, $P_{\mu e}$ is always 
small. Similarly, using Eq. (\ref{eq:tran3p2}) it is easy to 
see that $P_{\mu\tau}$ is large when $\sin^2\theta_{34}$ and 
$\sin^2\theta_{35}$ are large. In fact, its easy to see 
that non-zero $\sin^2\theta_{34}$ brings the $\mss$ driven 
first 
peak in $P_{\mu\tau}$ and non-zero 
$\sin^2\theta_{35}$ brings the $\msss$ driven second 
peak. 
Using similar arguments one can check that 
the muon neutrinos oscillate into the first sterile neutrino 
at the $\mss$ driven resonance and into the second 
sterile neutrino 
at the $\msss$ driven resonance. 

Since the mixing angle $\ses$ is non-zero, we expect resonant
transitions for electron neutrinos as well. We show in 
Fig. \ref{fig:3p2Eee} the oscillation probabilities 
associated with the electron type neutrinos and antineutrinos.
The upper left hand panel shows the survival probability
$P_{ee}$, the upper right hand panel shows 
$P_{e\tau}$, the lower 
left hand panel shows $P_{e\mu}$, while the 
lower right hand panel shows 
the transition probability to the first sterile state,
$P_{es_1}$ and to the second sterile state, $P_{es_2}$.
We have assumed the N2+N3 (or N2+I3) 
spectrum for the neutrinos with the 
current global best-fit numbers for the oscillation parameters. 
The black (dark) lines are for neutrinos while the cyan (light)
lines are for antineutrinos. For the I2+N3 (and I2+I3) spectra,
the black (dark) lines would be for antineutrinos and 
cyan (light) lines for neutrinos. 
As discussed before, we see that both resonances 
come in the neutrino (antineutrino) channel when 
$\mss>0$ and $\msss>0$ ($\mss<0$ and $\msss<0$). 
For the hybrid cases
H2+N3(a) and H2+N3(b) (as well as H2+I3(a) and H2+I3(b)),
only one resonance will occur in either the neutrino or 
the antineutrino channel depending whether the mass squared 
difference is positive or negative respectively. 

The expressions for the probabilities involving the 
electron neutrino in the 3+2 
picture using the parameterization for $U$ given by 
Eq. (\ref{eq:u3p2}) are
\be
P_{ee} \simeq 1 &-& \cos^4\theta_{15}^M\sin^22\theta_{14}^M  
\sin^2\bigg[\frac{(\Delta m_{41}^2)^M L}{4E}\bigg]
\nonumber\\
&-&
\cos^2\theta_{14}^M\sin^22\theta_{15}^M
\sin^2\bigg[\frac{(\Delta m_{51}^2)^M L}{4E}\bigg]
\nonumber\\
&-&
\sin^2\theta_{14}^M\sin^22\theta_{15}^M
\sin^2\bigg[\frac{(\Delta m_{54}^2)^M L}{4E}\bigg]\,.
\ee
The mixing angles $\sin^2\theta_{14}^M$ and $\sin^2\theta_{15}^M$ 
do not reach maximal value simultaneously. Therefore, we can 
see from this expression that we would have 2 big dips 
in the survival probability due to
the first and the second terms
when we have the 
$\mss$ and $\msss$ driven resonances respectively.
The last term is proportional to 
$\sin^2\theta_{14}^M\sin^22\theta_{15}^M$
and needs $\theta_{14}^M$ and and $\theta_{15}^M$ to be 
large simultaneously. A third dip would be possible 
only when this condition is satisfied.  
The transition probabilities have the 
general form given by Eq. (\ref{eq:tran3p2}).
The reason why $P_{e\mu} \simeq 0$ and 
$P_{e\tau} \simeq 0$ is same as that discussed before. 
If the 
$\nue  \rightarrow\nu_s$ resonance happens in the neutrino channel, 
the $\numu \rightarrow \nu_s$ resonance will happen in the antineutrino 
channel. Therefore, when $\sin^2\theta_{14}^M=1$ or 
$\sin^2\theta_{15}^M=1$ due to   $\nue \rightarrow \nu_s$ resonance, 
the other mixing angles do not receive any matter enhancement.
One can check that in this case the mixing matrix 
elements $U_{\mu 4}^M$,  $U_{\mu 5}^M$, $U_{\tau 4}^M$ and
$U_{\tau 5}^M$ are very small if the sterile mixing angles in 
vacuum are small and we have $P_{e\mu} \simeq 0$ and 
$P_{e\tau} \simeq 0$. 

\section{Flavor and Event Ratios with Sterile Neutrinos}

\begin{figure}[t]
\begin{center}
\includegraphics[width=16.0cm, height=9.0cm, angle=0]{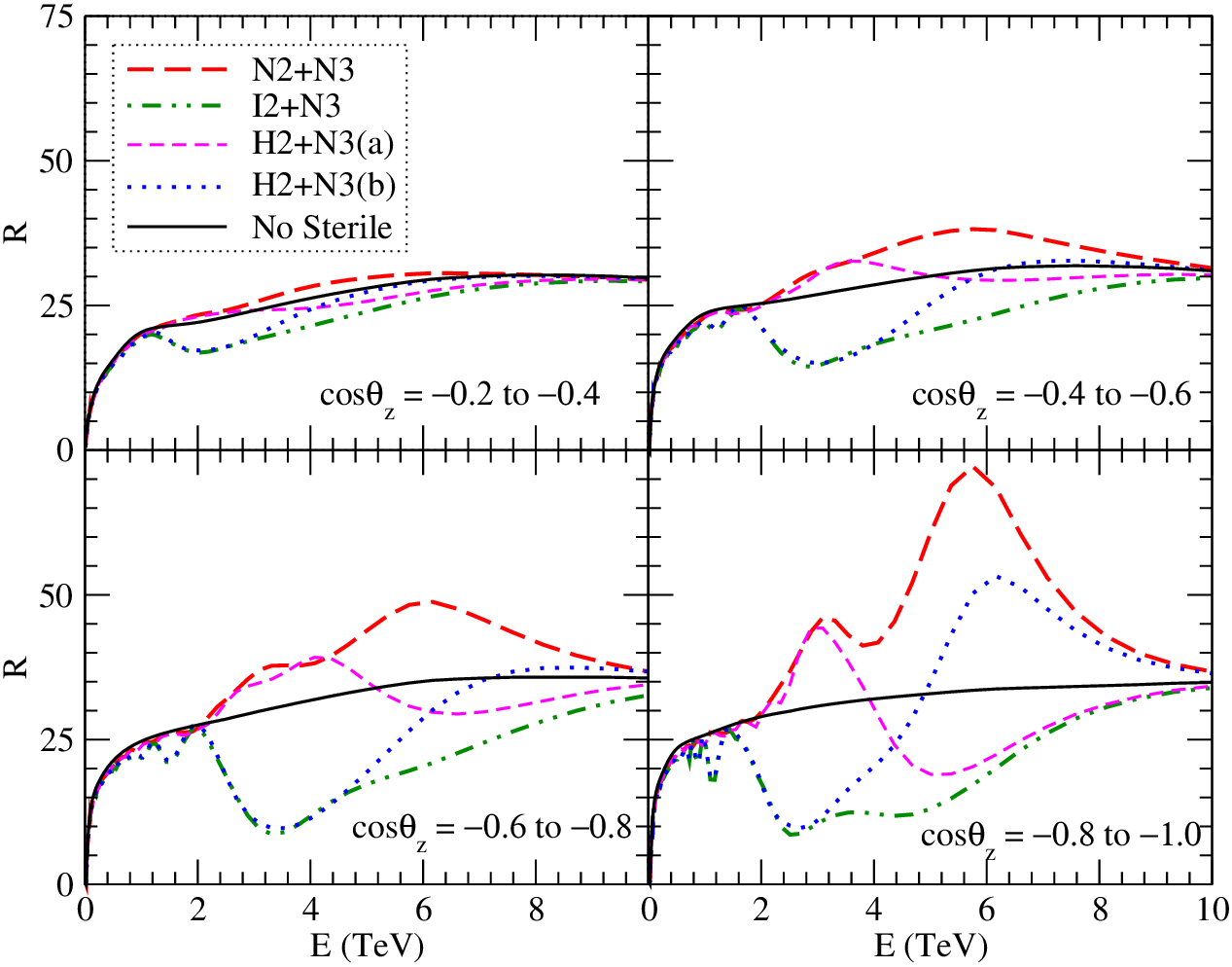}
\caption{\label{fig:rmu}
Binned result for the ratio $R$ as a function of energy $E$. 
The 
upper left panel shows the result for the zenith bin 
$-0.2 \geq \cos\theta_z \geq -0.4$, the 
upper right panel for 
$-0.4 \geq \cos\theta_z \geq -0.6$, the 
lower left panel for 
$-0.6 \geq \cos\theta_z \geq -0.8$, and the 
lower right panel for 
$-0.8 \geq \cos\theta_z \geq -1.0$. 
The solid lines show the expected $R$
in the respective zenith bins  
when we have no sterile neutrinos and 
there are only three generation oscillations. 
The other 4 line types 
correspond to  
N2+N3 (long dashed red lines),
I2+I3 (dot-dashed green lines), H2+N3(a) (dashed magenta lines)
and  H2+N3(b) (dotted blue lines). For all cases we have 
taken $|\mss| = 0.87$ eV$^2$, $|\msss| = 1.91$ eV$^2$ and 
mixing angles corresponding to their global best-fit values.
We take  $\sin^2\theta_{34} = \sin^2\theta_{35} 
=\sin^2\theta_{45}=0.01$. 
}
\end{center}
\end{figure}

Neutrino telescopes, such as IceCube and Km3Net are 
not expected to have any charge identification capability.
Therefore, they will not be able to distinguish the neutrino 
signal from the antineutrino signal. This is particularly 
relevant for matter effects, since resonant transitions due 
to Earth matter is large in only either the neutrino 
or the antineutrino channel for a given sign of the 
mass squared difference which drives the 
resonance. Consequently, one should look at the 
sum of the neutrino and antineutrino events expected at 
the neutrino telescope. 
The threshold energies for muon, electron and tau 
detection in IceCube are about 100 GeV, 1 TeV and 1000 TeV, 
respectively. 
It is easiest to see 
muon events, which leave 
distinct tracks in the detector. Both the electron
events and the tau events produce showers. 
In principle the tau events should be separable
from the electron events in certain energy range 
where they produce the so-called 
``double bang'' signal in the detector \cite{doublebang}.
However, because tau events have energy 
threshold of about 1000 TeV,  
they are not of any interest to us. 
Therefore, we could consider the simple event ratio 
\be
r = \frac{N_\numu + N_\anumu}
{N_\nue + N_\anue}~, 
\ee
where $N_\alpha $ are the number of events 
observed corresponding to the 
species type $\alpha$. 
The number of events is mainly given in terms of 
the (anti)neutrino flux, cross-section and the relevant oscillation 
probabilities. 
A detail prescription for calculating the number of 
events due to atmospheric 
$\numu$ is given in \cite{nutelnsi}.
In this paper, we will not attempt to calculate the 
number of events  exactly, which in addition to the main quantities 
mentioned above, also depend on other things like 
distance covered and  
energy loss of the lepton 
inside ice, details of the detector 
and the necessary cuts of the experiment.
Instead, just for the purpose of illustration, we 
present the ratio of the product of the 
flux, cross-section and 
the relevant probabilities as
\be
R = \frac{[\phi_\numu P_{\mu\mu}+\phi_\nue P_{e\mu}]\sigma_\nu + 
[\phi_\anumu P_{\bar\mu\bar\mu}+P_{\bar e\bar\mu}]\sigma_{\bar\nu}}
{[\phi_\nue P_{ee}+\phi_\numu P_{\mu e}]\sigma_\nu  + 
[\phi_\anue P_{\bar e \bar e}+
\phi_\anumu P_{\bar \mu \bar e}]\sigma_{\bar\nu}
}~.  
\label{eq:ratioapprox}
\ee
We use the atmospheric neutrino flux given by Honda {\it et. al.}
\cite{honda} and high energy charged current 
cross-sections from \cite{uhecross}. 
The efficiency of observing electron events is smaller than 
for muon events. However the difference in the $1-10$ TeV 
range is small and therefore we neglect that here, since 
what we present is merely for illustration only. 
IceCube is expected to 
have rather good zenith angle resolution of about $25^\circ$
\cite{uheflavor} and in \cite{nutelnsi} the authors have 
presented their results in 5 energy bins between $1-10$ TeV. 
In Fig. \ref{fig:rmu} we show the zenith angle binned
value for $R$, as a function of 
the energy $E$. 
We have divided the zenith angle range 
$-0.2 \geq \cos\theta_z \geq -1.0$, into 4 bins and 
show results where we have calculated $R$ by 
summing over the product of the flux, cross-section and 
relevant probabilities in the zenith bins. The 
upper left panel shows the result for the zenith bin 
$-0.2 \geq \cos\theta_z \geq -0.4$, the 
upper right panel for 
$-0.4 \geq \cos\theta_z \geq -0.6$, the 
lower left panel for 
$-0.6 \geq \cos\theta_z \geq -0.8$, and the 
lower right panel for 
$-0.8 \geq \cos\theta_z \geq -1.0$. 
The solid black lines show the expected $R$
in the respective zenith bins  
when we have no sterile neutrinos and 
there are only three generation oscillations. Note that 
these three generation oscillations are important for 
neutrino energies up to 1 TeV for neutrinos travelling 
large distances inside Earth. The other 4 line types 
correspond to the 4 relevant mass spectra 
discussed for the 3+2 scheme, 
N2+N3 (long dashed red lines),
I2+I3 (dot-dashed green lines), H2+N3(a) (dashed magenta lines)
and  H2+N3(b) (dotted blue lines). For all cases we have 
taken $|\mss| = 0.87$ eV$^2$, $|\msss| = 1.91$ eV$^2$ and 
mixing angles corresponding to their global best-fit values,
as discussed in Section 4. The hitherto unconstrained mixing 
angles $\sin^2\theta_{34}$, $\sin^2\theta_{35}$ and 
$\sin^2\theta_{45}$ are taken as 0.01.   
We see that there is a huge change in the 
value of $R$ due to presence of sterile neutrinos for 
all values of $E$ between 1 and 10 TeV. The change is also 
seen to be clearly dependent on the energy and 
zenith angle of the neutrinos. 
We note from the figure that not only should it be possible 
to establish the 
presence of 
sterile neutrinos from the observations,
it should also be possible 
to differentiate between the different 3+2 neutrino mass
because each one of them has a distinct prediction for 
$R$. 

Few comments are in order. 
The number of muon events expected from high energy atmospheric 
neutrinos has been given in Table 1 of \cite{nutelnsi}, 
for 2 zenith angle bins $-0.6 \geq \cos\theta_z \geq -1.0$
and $-0.2 \geq \cos\theta_z \geq -0.6$ and different 
energy bins. In the $-0.6 \geq \cos\theta_z \geq -1.0$ bin, 
about 52,474 muon events are expected in 
$0.1 \leq E~({\rm TeV}) \leq 0.16$ energy bin, 
about 3,330 events in 
$0.25 \leq E~({\rm TeV}) \leq 3.98$ energy bin and
about 1,721 events in  
$3.98 \leq E~({\rm TeV}) \leq 6.31$ energy bin, after 
10 years of IceCube operation. 
Higher number of events are expected in the 
$-0.2 \geq \cos\theta_z \geq -0.6$ zenith angle bin. 
Thus we expect a 
rather good statistic atmospheric neutrino data 
at the neutrino telescopes. 
In fact, with such high statistic, it should be possible 
to look for sterile neutrinos in the atmospheric neutrino 
data sample using just the muon events alone.   


\section{Conclusions}

Following the recently declared MiniBooNE results, 
sterile neutrinos have been the focus of discussions in 
the field of neutrino physics. In particular, the question 
whether MiniBooNE data has unambiguously ruled out the 
possibility of sterile neutrinos needed to explain the 
LSND results has been raised. In this paper we have 
expounded the possibility of answering this question 
using the high energy atmospheric neutrino data in the 
upcoming neutrino telescopes. 

If sterile neutrinos exist with $\dm \sim $ eV$^2$, we 
expect to see flavor oscillations of upward going atmospheric 
neutrinos with the peak in the 
transition probability at an energy of a few TeV. One 
could naively think that these oscillations would normally be 
small owing to the smallness of the sterile mixing angles, 
which are severely constrained by the short baseline 
oscillation experiment data.  
We pointed out that near-resonant matter effects driven by the 
sterile neutrino mass squared differences drive these very small
mixing angles in vacuum to almost maximal in matter. 
For a given neutrino baseline inside the Earth, 
the largest oscillations of course occur when the condition of 
resonance and the condition of oscillation peak are simultaneously
satisfied. We showed in the framework of the simpler 3+1 neutrino 
mass scenario that this condition could be satisfied for 
TeV neutrinos crossing the Earth. 
We assumed a simple framework where the only non-zero sterile 
angle was $\theta_{24}$ and 
calculated the value of the 
mixing angle for which one could have maximal $\numu$ oscillations
into sterile neutrino.
For $\theta_{24}$ both larger and smaller than this critical value, 
one would get lesser oscillation. We next allowed for non-zero 
$\theta_{34}$ values and studied how this mixing angle 
changed the oscillation scenario. We showed that matter effects 
simultaneously enhance both $\theta_{24}$ and $\theta_{34}$ and 
we have genuine three generation effects in the oscillation 
probability. One very important effect is that with non-zero 
$\theta_{34}$ the $P_{\mu\tau}$ oscillation probability 
increases significantly and could even become maximal, while 
the $P_{\mu s}$ probability simultaneously decreases. 

We considered the still viable 3+2 neutrino mass and mixing scheme 
and presented the oscillation probabilities when all 
mixing angles were allowed to be non-zero, as needed to explain the 
global oscillation data, including LSND and MiniBooNE.
For the 3+2 mass scheme one can have as many as 8 different 
mass ordering. Of these, there are at least 4 different 
possibilities that would allow for resonant 
matter transition driven by the sterile mass eigenstates and
each one gives a distinct signature in the oscillation 
pattern.  
We presented the results for the oscillation probabilities 
obtained by evolving the full five generation neutrino 
system inside the Earth matter as they travel, assuming the
PREM profile for the matter density. We  
explained these results 
using simplified constant matter density picture. 
We showed that the mixing angles $\theta_{34}$ and $\theta_{35}$
which are enhanced inside the Earth matter cause 
$P_{\mu\tau}$ to increase significantly. 
We emphasized the fact that while $\numu \rightarrow \nu_s$
resonance occurs for $\dm <0$ in the neutrino channel, 
the $\nue  \rightarrow \nu_s$ resonance condition 
is satisfied for  $\dm > 0$ in the neutrino channel. 
In the antineutrino channel of course the sign of $\dm$ 
is reversed for the resonance condition to be satisfied. 
We showed how this feature ensured that the transition 
probability $P_{e \mu}$ and $P_{e \tau}$ always remained 
negligible. For $\theta_{14}$ and/or $\theta_{15}$ non-zero, 
we showed how resonance in the neutrino (antineutrino) channel 
produces huge dips in $P_{ee}$. 

Finally, we discussed how these large matter effects due to 
presence of sterile neutrinos would show up in the 
neutrino telescopes. Atmospheric neutrinos form a ``background'' 
for the ultra high energy neutrino observation in the neutrino 
telescopes. These atmospheric neutrinos are in the TeV range 
where we expect near-resonant matter effects. At these energies  
hundreds of thousands 
of muon type events are expected from atmospheric 
neutrinos in IceCube and we 
argued that even a very moderate energy and zenith angle resolution 
in the data would lead to an unambiguous signal for sterile neutrinos.

\vskip 1.0cm
{\Large \bf Note Added}
\vglue 0.3cm
\noindent
After the first version of the this paper appeared, 
the Super-Kamiokande collaboration 
have released their data on showering muon events, which 
come from neutrinos with energies in the TeV 
range \cite{sktev}. 
They comment on the feasibility of 
using their very high energy upward-going muon data sample
which come from showing type events to shed light on the 
existence of high $\Delta m^2$ solutions, and cite this paper.



\end{document}